\documentclass[pra,preprint,floatfix,longbibliography]{revtex4-1}
\usepackage{hyperref} 
\usepackage{graphicx} 
\usepackage{subfig}
\usepackage{color}
\usepackage{filecontents}
\usepackage{soul}

\newcommand{\ket}[1]{\left| #1 \right\rangle}

\usepackage{dcolumn}
\usepackage{bm}
\usepackage{amssymb,amsmath}
\usepackage{color}
\usepackage{float}
\usepackage{tikz}
\usetikzlibrary{arrows}
\begin{document}

\title{Dimer chains in waveguide quantum electrodynamics}

\author{Imran M. Mirza}
\affiliation{Department of Mathematics, University of Michigan, Ann Arbor, Michigan 48109, USA}
\email{imranmir@umich.edu}

\author{Jeremy G. Hoskins}
\affiliation{Department of Mathematics, Yale University, New Haven, CT 06511, USA}
\email{jeremy.hoskins@yale.edu}

\author{John C. Schotland}
\affiliation{Department of Mathematics and Department of Physics, University of Michigan, Ann Arbor, Michigan 48109, USA}
\email{schotland@umich.edu}

\begin{abstract}
We examine the propagation of single photons in periodic and disordered dimer chains coupled to one-dimensional chiral and bidirectional waveguides. Each dimer is composed of two dipole-coupled atoms. In the disordered setting, we separately treat two types of position disorder, namely in dimer length and in dimer separation. The focus of this study is to understand in what ways the interplay between dipole-dipole interactions and directionality of photon  emission can impact the transport of single photons. Cold atoms trapped near optical fibers can serve as an experimentally realizable platform for the models that we consider.

\end{abstract}

\maketitle

\section{Introduction}
Multiatom waveguide quantum electrodynamics (QED) is a powerful platform to demonstrate a range of novel quantum optical effects~\cite{roy2017colloquium}. These include: strong light-matter interactions \cite{hung2013trapped,*javadi2015single,*yalla2014cavity}, electromagnetically induced transparency (EIT) and dark state formation \cite{martens2013photon,*chang2014quantum}, bunching and antibunching of photons \cite{fan2010input, *xu2017input}, photonic band gaps and slow light \cite{shen2007stopping, *douglas2015quantum}, localization of photons \cite{witthaut2010photon, mirza2017chirality, mirza2017influence}, resonance fluorescence \cite{kocabacs2012resonance}, multi-atom entangled states \cite{gonzalez2015chiral, mirza2016multiqubit, mirza2016two}, chiral photonic emission \cite{coles2016chirality, *mitsch2014quantum}, single photon switches \cite{tiecke2014nanophotonic}, and quantum gates \cite{paulisch2016universal}, among others. All of these effects have applications to understanding light-matter interactions at the quantum level and also to building novel quantum technologies.
 
The recent demonstration of the enhancement of optical spin-orbit coupling due to confinement of light in subwavelength structures has led to the emergence of the field of {\it chiral quantum optics} \cite{lodahl2017chiral}. Waveguide QED is closely linked with this development. For instance, it has been shown that up to 90\% chiral photon emission into waveguide modes is achievable in practice~\cite{coles2016chirality}. In addition, we and others have shown that chirality can be used to enhance multiqubit entanglement \cite{gonzalez2015chiral, mirza2016multiqubit, mirza2016two} and also influences photon transport in the presence of disorder \cite{mirza2017chirality,mirza2017influence}. Some of the key findings of our work are that photon transmission in chiral waveguides (coupled to either two-level or three-level atoms) is immune to position disorder. We have also shown that single photon localization occurs in chiral waveguides with disordered atomic transition frequencies. Moreover, localization also occurs in bidirectional waveguides with both frequency and position disorder.

In waveguide QED, varying the atomic separation can produce a variety of physical effects. If the separation is larger than half of the wavelength of the optical field, then non-Markovian effects become important \cite{milonni1974retardation}. On the other hand, if the interatomic separation becomes much smaller than the wavelength, then collective effects become important, and can lead to superradiance and subradiance \cite{scully2009super, roof2016observation}. In addition, small interatomic separation also plays a key role in the phenomena of quantum beats \cite{zheng2013persistent}, entanglement evolution \cite{facchi2016bound} and Bragg mirrors \cite{corzo2016large}. We also note that if the interatomic separation becomes considerably smaller than the wavelength at resonance, then interatomic dipole-dipole interactions must be taken into account \cite{agarwal2013quantum}. In this context, Cheng et al. \cite{cheng2017waveguide} have shown that Fano interference can be used to estimate the strength of such interactions. In addition, they have found that in periodically arranged atoms bidirectionally coupled to a waveguide, 
photon transport is extremely sensitive to the atomic positions. 
This sensitivity originates from the dependence of the dipole-dipole interaction on the separation between the atoms. Hence, small variations in the atomic positions can modify the strength of the interaction considerably \cite{marcuzzi2017facilitation}. 

Motivated by the above considerations, in this paper we consider a model for waveguide QED involving pairs of atoms (dimers), where the atoms comprising the dimer are strongly coupled,  but the dimers themselves are sufficiently far apart so that interdimer interactions can be neglected. In this setting, 
we investigate the transmission of single photons in periodic and disordered chains of dimers. We consider two types of disorder: random dimer separations and random dimer lengths. Our results can be summarized as follows. For chiral waveguides, we find that in the periodic case, the presence of dipole-dipole interactions leads to a splitting in the frequency-dependent transmission curve. In the disordered case, we find immunity to disorder for random dimer separations. We also provide numerical evidence of localization for random dimer lengths. The latter result should be contrasted with the case of chiral waveguide QED, where localization does not occur with position disorder~\cite{mirza2017chirality}. For symmetric bidirectional waveguides, in the periodic case, we find modifications of the band structure due to interactions. Localization occurs for both types of disorder. Moreover, the localization length becomes larger in the presence of interactions. 

There has been considerable recent experimental progress in driven-dissipative Rydberg trapped atoms \cite{vermersch2016implementation, glaetzle2014quantum, hofmann2014experimental}, nitrogen-vacancy centers \cite{fu2008coupling}, ultracold atoms  coupled to waveguide resonators (including superconducting waveguides) \cite{hattermann2017coupling,lalumiere2013input}. Thus the model presented in this paper would seem to be not far from physical realization. One possible implementation consists of two interpenetrating lattices of two-level atoms, where the atoms belonging to different sublattices are close enough to form dimers. Since small deviations from a perfectly ordered lattice can change the strength of dipole-dipole interactions, the results of this paper are potentially important for understanding photon transport in realistic dimer systems.

This paper is organized as follows. In Section II we introduce the model  system under consideration. In Sections III and IV, we address the problems of periodic and disordered chiral and bidirectional waveguides, respectively. Finally, in Section V, we conclude with a discussion of our results.
  
\begin{figure}[t]
\centering
\includegraphics[width=6.8in, height=2in]{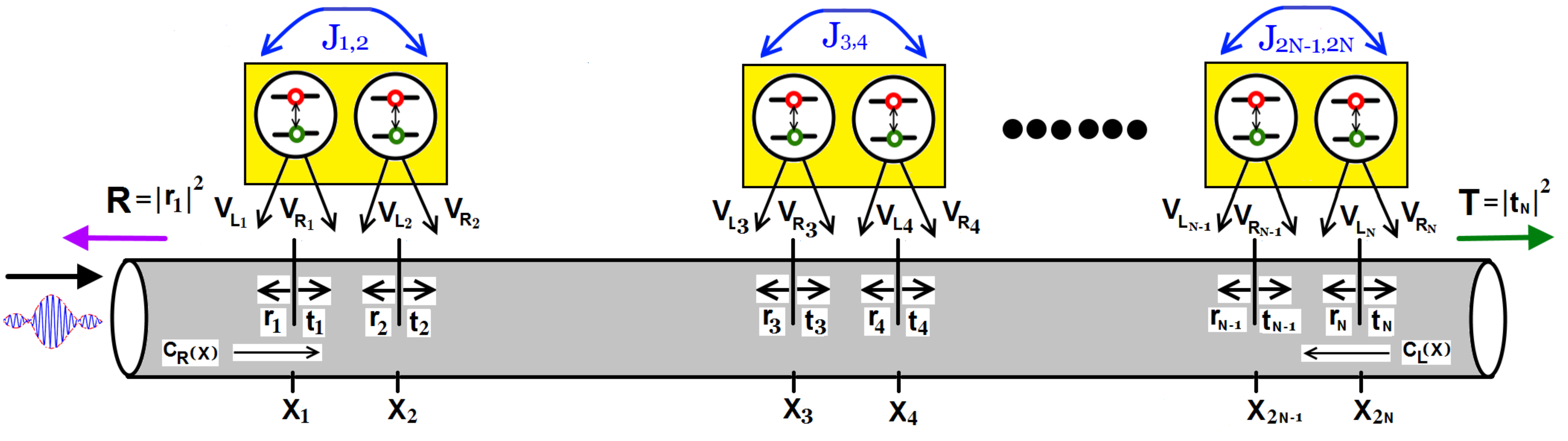} 
\captionsetup{
  format=plain,
  margin=1em,
  justification=raggedright,
  singlelinecheck=false
}
\caption{(Color online) Illustrating the multidimer waveguide QED system considered in this paper.}
\label{Fig1}
\end{figure}

\section{Model System}

We consider a system of $N$ atomic dimers coupled to a waveguide. Each dimer is composed of two identical atoms. We note that the separation of the atoms within a dimer and the spacing between dimers can be varied, as illustrated in Fig.~\ref{Fig1}. The Hamiltonian of the system is of the form
\begin{equation}
\label{eq:Hsys}
\begin{split}
\hat{H}=&\hbar\sum_{j}(\Omega-i\gamma)\hat{\sigma}^{\dagger}_{j}\hat{\sigma}_{j}+\hbar\sum^{2N-1}_{j=1,3}\Bigg(J_{j,j+1}\hat{\sigma}^{\dagger}_{j}\hat{\sigma}_{j+1}+J_{j+1,j}\hat{\sigma}^{\dagger}_{j+1}\hat{\sigma}_{j}\Bigg) \\
&+\hbar\int dx\hat{c}^{\dagger}_{R}(x)\left(\omega_{0} - iv_{R}\frac{\partial}{\partial x}\right)\hat{c}_{R}(x)  + \hbar\int dx\hat{c}^{\dagger}_{L}(x)\left(\omega_{0} + iv_{L}\frac{\partial}{\partial x}\right)\hat{c}_{L}(x)\\
&+ \hbar\sum_{m}\sum^{2N}_{j=1}\int dx\delta(x-x_{j})\left[{V_{m}}\hat{c}^{\dagger}_{m}(x)\hat{\sigma}_{j}+h.c.\right] , 
\end{split}
\end{equation}
where we have made the rotating wave approximation and employ real-space quantization of the optical field~\cite{bozhevolnyi2017quantum}. The first term is the Hamiltonian of the atoms. The lowering operator for atom $j$ is denoted $\hat{\sigma}_{j}$ and the corresponding raising operator is $\hat{\sigma}_{j}^\dag$. The atomic resonance energy is $\hbar\Omega$ and the rate of spontaneous emission into nonwaveguide modes is accounted for by the term $i\gamma$. Note that the atomic operators obey the anticommutation relations $\lbrace\hat{\sigma}_{i},\hat\sigma_{j}^\dag \rbrace=\delta_{ij}$.

The second term in the Hamiltonian accounts for dipole-dipole interactions within each dimer. Here the coupling $J_{i,j}$ between the $i$th and $j$th atoms is given by~\cite{agarwal2013quantum,cheng2017waveguide}
\begin{equation}
J_{i,j}=\frac{3\Gamma_{0}}{4}\Bigg( \frac{\cos x_{ij}}{x^3_{ij}}+\frac{\sin x_{ij}}{x^2_{ij}}-\frac{\cos x_{ij}}{x_{ij}}\Bigg) ,
\end{equation}
where $\Gamma_{0}$ is the free-space atomic decay rate, {$x_{ij} =\Omega|x_{i}-x_{j}|/c$}, and $x_i$ is the position of the $i$th atom. For simplicity, we also assume that the dipole moments of the atoms are perpendicular to the axis of the waveguide.

The third and fourth terms in the Hamiltonian comprise the Hamiltonian of the waveguide. The waveguide dispersion relation has been linearized about the frequency $\omega_0$. In general, the waveguide supports right- and left going modes with group velocities $v_R$ and $v_L$, respectively. The destruction of a photon at position $x$ in the left (right) waveguide continuum is represented by the field operator $\hat{c}_{L}(x)(\hat{c}_{R}(x))$. Likewise, the creation of a photon at position $x$ in the left (right) waveguide continuum is represented by $\hat{c}_{L}^\dag(x)(\hat{c}_{R}^\dag(x))$. The nonvanishing commutation relations between the field operators are of the form
\begin{eqnarray}
\left[\hat{c}_m(x),\hat{c}^\dag_{n}(x')\right] = \delta_{mn}\delta(x-x') ,\end{eqnarray}
where $m\in\{R,L\}$.

The final term in the Hamiltonian describes the atom-field interaction. Here the coupling $V_{m}$ is taken to be real-valued. The case of a symmetric waveguide corresponds to $v_R=v_L$ and $V_{R}=V_{L}$. The extension to non-symmetric waveguides is straightforward, but will not be directly considered. If either $v_R$ or $v_L$ is set to zero, then the above model describes a chiral waveguide.

The single-photon eigenstate of $\hat H$ is of the form
\begin{equation}
\label{eq:Psi}
\begin{split}
&\ket{\Psi}=\sum_{m}\int dx \varphi_{m}(x)\hat{c}^{\dagger}_{m}(x)\ket{\varnothing}+\sum_{j}a_{j}\hat{\sigma}^{\dagger}_{j}\ket{\varnothing}.
\end{split}
\end{equation} 
Here $\varphi_{R}(x)$ $(\varphi_{L}(x)$) is the amplitude of finding the photon in the right (left) waveguide continuum. The quantity $a_{j}$ is the amplitude for the $j$th atom to be in its excited state and $\ket{\varnothing}$ is the overall ground state of the system, where no photons are present in the waveguide and all atoms are in their ground state. Making use of the time-independent Schr\"odinger equation $\hat{H}\ket{\Psi}=\hbar\omega\ket{\Psi}$ along with (\ref{eq:Hsys}) and (\ref{eq:Psi}), we find that the amplitudes obey the equations
\begin{subequations}
\label{eq:TIAENC}
\begin{eqnarray}
-iv_{R}\frac{\partial \varphi_{R}(x)}{\partial x}+\sum^{N}_{j=1,2}V_{R}\delta(x-x_{j})a_{j}=(\omega-\omega_0)\varphi_{R}(x) , \\ 
iv_{L}\frac{\partial \varphi_{L}(x)}{\partial x}+\sum^{N}_{j=1,2}V_{L}\delta(x-x_{j})a_{j}=(\omega-\omega_0)\varphi_{L}(x) , \\ 
V_{R}\varphi_{R}(x_{j})+V_{L}\varphi_{L}(x_{j})=(\omega-\Omega+i\gamma)a_{j}-J_{j,j+1}a_{j+1},\label{eq:phiNC1} \\
V_{R}\varphi_{R}(x_{j+1})+V_{L}\varphi_{L}(x_{j+1})=(\omega-\Omega+i\gamma)a_{j+1}-J_{j+1,j}a_{j} \label{eq:phiNC2},
\end{eqnarray}
\end{subequations}
where in (\ref{eq:phiNC1}) and (\ref{eq:phiNC2}) $j$ is odd and $\omega$ is the incoming single-photon frequency. Elimination of the atomic amplitudes from the above equations yields
\begin{subequations}
\label{eq:phiEq}
\begin{eqnarray}
-iv_{R}\frac{\partial \varphi_{R}(x)}{\partial x}+\sum^{2N-1}_{j=1,3}\Bigg[\Bigg(\frac{\delta\omega\delta(x-x_{j}) V_{R}}{\delta\omega^2-J^{2}_{j,j+1}} \Bigg)(V_{R}\varphi_{R}(x)+V_{L}\varphi_{L}(x) )+\Bigg(\frac{J_{j,j+1}V_{R}\delta(x-x_{j+1})}{\delta\omega^2-J^{2}_{j,j+1}}\Bigg)\nonumber\\
\times(V_{R}\varphi_{R}(x_{j})+V_{L}\varphi_{L}(x_{j})) \Bigg]+\sum^{2N}_{j=2,4}\Bigg[\Bigg(\frac{\delta\omega\delta(x-x_{j}) V_{R}}{\delta\omega^2-J^{2}_{j-1,j}} \Bigg)(V_{R}\varphi_{R}(x)+V_{L}\varphi_{L}(x_{j}) )\nonumber\\
+\Bigg(\frac{J_{j-1,j}V_{R}\delta(x-x_{j-1})}{\delta\omega^2-J^{2}_{j-1,j}}\Bigg)(V_{R}\varphi_{R}(x_{j})+V_{L}\varphi_{L}(x_{j})) \Bigg]=(\omega-\omega_{0})\varphi_{R}(x),\hspace{10mm}\\ 
 \nonumber \\
iv_{L}\frac{\partial \varphi_{L}(x)}{\partial x}+\sum^{2N-1}_{j=1,3}\Bigg[\Bigg(\frac{\delta\omega\delta(x-x_{j}) V_{L}}{\delta\omega^2-J^{2}_{j,j+1}} \Bigg)(V_{R}\varphi_{R}(x)+V_{L}\varphi_{L}(x) )+\Bigg(\frac{J_{j,j+1}V_{L}\delta(x-x_{j+1})}{\delta\omega^2-J^{2}_{j,j+1}}\Bigg)\nonumber\\
\times(V_{R}\varphi_{R}(x_{j})+V_{L}\varphi_{L}(x_{j})) \Bigg]+\sum^{2N}_{j=2,4}\Bigg[\Bigg(\frac{\delta\omega\delta(x-x_{j}) V_{L}}{\delta\omega^2-J^{2}_{j-1,j}} \Bigg)(V_{R}\varphi_{R}(x)+V_{L}\varphi_{L}(x_{j}) )\nonumber\\
+\Bigg(\frac{J_{j-1,j}V_{L}\delta(x-x_{j-1})}{\delta\omega^2-J^{2}_{j-1,j}}\Bigg)(V_{R}\varphi_{R}(x_{j})+V_{L}\varphi_{L}(x_{j})) \Bigg]=(\omega-\omega_{0})\varphi_{L}(x),\hspace{10mm}
\end{eqnarray}
\end{subequations}
where $\delta\omega=\omega-\Omega-i\gamma$. We note that the presence of dipole-dipole interactions splits the above equation into even and odd terms. 

To construct the solution of (\ref{eq:phiEq}), we observe that the single photon wavefunctions between any two dimers take the form $\varphi_{R}(x)\propto e^{iq_{R}x}$ and $\varphi_{L}(x)\propto e^{-iq_{L}x}$, where the wavenumbers associated with the right and left field amplitudes are defined by $q_{R}=(\omega-\omega_{0})/v_{R}$ and $q_{L}=(\omega-\omega_{0})/v_{L}$, respectively. Therefore, we write
\begin{equation}
\label{eq:phiNCR}
\varphi_{R}(x)=
\begin{cases}
  e^{iq_{R}x}, \hspace{5mm}x<x_1,\\      
  t_{1}e^{iq_{R}x}, \hspace{2mm} x_{1} \le x \le x_2, \\
  \vdots \\
  t_{N}e^{iq_{R}x}, \hspace{2mm}x>x_N  .
\end{cases}
\end{equation}
and
\begin{equation}
\label{eq:phiNCL}
\varphi_{L}(x)=
\begin{cases}
  r_{1}e^{-iq_{L}x}, \hspace{5mm}x<x_1,\\      
  r_{2}e^{-iq_{L}x}, \hspace{2mm} x_{1} \le x \le x_2, \\
  \vdots \\
   r_{N}e^{-iq_{L}x}, \hspace{2mm} x_{N-1} \le x \le x_N, \\
  0, \hspace{2mm}x>x_N  .
\end{cases}
\end{equation}
where $r_{N+1}=0$ and $t_{0}=1$. The coefficients $t_{j}$ and $r_{j}$ are obtained by integrating (\ref{eq:phiEq}) over the interval $[x_{j}-\epsilon,x_{j}+\epsilon]$, which yields the following jump conditions
\begin{subequations}
\begin{eqnarray}
iv_{R}\left[\varphi_{R}(x_{j}+\epsilon)-\varphi_{R}(x_{j}-\epsilon)\right]=\delta\omega\Bigg(\frac{V_{R}}{\delta\omega^2-J^{2}_{j,j+1}}\Bigg)(V_{R}\varphi_{R}(x_{j})+V_{L}\varphi_{L}(x_{j}))\nonumber\\
+J_{j,j+1}\Bigg(\frac{V_{R}}{\delta\omega^2-J^{2}_{j,j+1}}\Bigg)(V_{R}\varphi_{R}(x_{j})+V_{L}\varphi_{L}(x_{j})),\hspace{5mm}\textit{j \rm odd,}\hspace{2mm}\\
iv_{R}\left[\varphi_{R}(x_{j}+\epsilon)-\varphi_{R}(x_{j}-\epsilon)\right]=\delta\omega\Bigg(\frac{V_{R}}{\delta\omega^2-J^{2}_{j-1,j}}\Bigg)(V_{R}\varphi_{R}(x_{j})+V_{L}\varphi_{L}(x_{j}))\nonumber\\
+J_{j-1,j}\Bigg(\frac{V_{R}}{\delta\omega^2-J^{2}_{j-1,j}}\Bigg)(V_{R}\varphi_{R}(x_{j-1})+V_{L}\varphi_{L}(x_{j-1})),\hspace{5mm}\textit{j \rm even,}\hspace{2mm}\\
iv_{L}\left[\varphi_{L}(x_{j}+\epsilon)-\varphi_{L}(x_{j}-\epsilon)\right]=-\delta\omega\Bigg(\frac{V_{L}}{\delta\omega^2-J^{2}_{j,j+1}}\Bigg)(V_{R}\varphi_{R}(x_{j})+V_{L}\varphi_{L}(x_{j}))\nonumber\\
-J_{j,j+1}\Bigg(\frac{V_{L}}{\delta\omega^2-J^{2}_{j,j+1}}\Bigg)(V_{R}\varphi_{R}(x_{j})+V_{L}\varphi_{L}(x_{j})),\hspace{5mm}\textit{j \rm odd,}\hspace{2mm}\\
iv_{L}\left[\varphi_{L}(x_{j}+\epsilon)-\varphi_{L}(x_{j}-\epsilon)\right]=-\delta\omega\Bigg(\frac{V_{L}}{\delta\omega^2-J^{2}_{j-1,j}}\Bigg)(V_{R}\varphi_{R}(x_{j})+V_{L}\varphi_{L}(x_{j}))\nonumber\\
-J_{j-1,j}\Bigg(\frac{V_{L}}{\delta\omega^2-J^{2}_{j-1,j}}\Bigg)(V_{R}\varphi_{R}(x_{j-1})+V_{L}\varphi_{L}(x_{j-1})),\hspace{5mm}\textit{j \rm even.}\hspace{2mm}
\end{eqnarray}
\end{subequations}
Next, introducing the quantities $\Gamma_{R}=V_{R}^{2}/2v_{R}$ and $\Gamma_{L}={V_{L}^{2}}/{2v_{L}}$ and regularizing the discontinuity in $\varphi_{m}$ by 
\begin{equation}
\varphi_{m}(x)=\lim_{\epsilon\longrightarrow 0}\left[\varphi_{m}(x_{j}+\epsilon)+\varphi_{m}(x_{j}-\epsilon)\right]/2, 
\end{equation}
and using (\ref{eq:phiNCR}) and (\ref{eq:phiNCL}), we obtain the recursion relations
\begin{equation}
\label{eq:recurNC}
\begin{split}
& t_{j}-t_{j-1}= -i\alpha^{(R)}_{j,j+1}(t_{j}+t_{j-1})-i\alpha^{(RL)}_{j,j+1}(r_{j+1}+r_{j})-i\beta^{(R)}_{j,j+1}e^{i\theta_{j,j+1}}(t_{j}+t_{j+1})\\
&-i\beta^{(RL)}_{j,j+1}e^{i\theta_{j,j+1}}(r_{j+1}+r_{j}),\\
&  t_{j}-t_{j-1}=-i\alpha^{(R)}_{j-1,j}(t_{j}+t_{j-1})-i\alpha^{(RL)}_{j-1,j}(r_{j+1}+r_{j})-i\beta^{(R)}_{j-1,j}e^{i\theta_{j-1,j}}(t_{j-1}+t_{j-2})\\
&-i\beta^{(RL)}_{j-1,j}e^{i\theta_{j-1,j+1}}(r_{j}+r_{j-1}),\\
& r_{j+1}-r_{j}=i\alpha^{(LR)}_{j,j+1}(t_{j}+t_{j-1})+i\alpha^{(L)}_{j,j+1}(r_{j+1}+r_{j})+i\beta^{(LR)}_{j,j+1}e^{-i\theta_{j,j+1}}(t_{j}+t_{j-1})\\
&+i\beta^{(L)}_{j,j+1}e^{-i\theta_{j,j+1}}(r_{j+1}+r_{j}),\\
& r_{j+1}-r_{j}=i\alpha^{(LR)}_{j-1,j}(t_{j}+t_{j-1})+i\alpha^{(L)}_{j-1,j}(r_{j+1}+r_{j})+i\beta^{(LR)}_{j-1,j}e^{-i\theta_{j-1,j}}(t_{j-1}+t_{j-2})\\
&+i\beta^{(L)}_{j-1,j}e^{-i\theta_{j-1,j}}(r_{j}+r_{j-1}).
\end{split}
\end{equation}
Here
\begin{equation}
\begin{split}
&\alpha^{(R/L)}_{j,j+1} = \frac{\Gamma_{R/L}\delta\omega}{\delta\omega^2-J^{2}_{j,j+1}}, \quad \alpha^{(RL/LR)}_{j,j+1}=\sqrt{\frac{v_{L/R}}{v_{R/L}}}\frac{\sqrt{\Gamma_{R}\Gamma_{L}}\delta\omega}{\delta\omega^2-J^{2}_{j,j+1}}, \\
&\beta^{(L/R)}_{j,j+1}=\frac{J_{j,j+1}\Gamma_{L/R}}{\delta\omega^2-J^{2}_{j,j+1}}, \quad \beta^{(RL/LR)}_{j,j+1} = \sqrt{\frac{v_{L/R}}{v_{R/L}}}\frac{J_{j,j+1}\sqrt{\Gamma_{R}\Gamma_{L}}}{\delta\omega^2 -J^{2}_{j,j+1}},
\end{split}
\end{equation}
$\theta_{j,j+1}=2\pi (x_{j}-x_{j+1})/\lambda_{QD}$, where $\lambda_{QD}$ is the wavelength corresponding to the transition frequency of the emitter. Note that $j$ is odd in the first two equations of (\ref{eq:recurNC}) and even in the last two equations. The transmission and reflection coefficients can be expressed in terms of the phase accumulated by the photon while traveling between two consecutive dimers through the waveguide
according to
\begin{equation}
t_{j}=\widetilde{t}_{j}e^{-i(q_{R}+q_{L})x_{j}/2},\hspace{5mm} r_{j}=\widetilde{r}_{j}e^{i(q_{R}+q_{L})x_{j-1}/2}.
\end{equation}
After some rearrangement, (\ref{eq:recurNC}) can be expressed in the form of the matrix recursion relation
\begin{equation}
\label{eq:TransEq}
\begin{pmatrix}
    \widetilde{t}_{2j}     \\
    \widetilde{r}_{2j-1}  \\
\end{pmatrix}=\mathcal{T}_{j}\begin{pmatrix}
\widetilde{t}_{2j-2}\\
\widetilde{r}_{2j+1}\\
\end{pmatrix}.
\end{equation} 
Here $\mathcal{T}_{j}$ is the transfer matrix which describes the input and output fields from the $j$th dimer in the array. The form of $\mathcal{T}_{j}$ is rather involved and is not presented here. The net transfer matrix $M$ for an $n$ dimer system is given by
\begin{equation}
M=\prod_{j}\mathcal{T}_j
\end{equation}
The net transmission coefficient from a chain of $n$ dimers is given by $T=|t_{2N}|^{2}$ while $n=2N$ and the reflection coefficient is $R=|r_{1}|^{2}$.

\section{Chiral waveguides}
For a chiral waveguide with only a right-going mode we set $\Gamma_{L}=0$ and $\Gamma_{R}=\Gamma$. Then, making use of (\ref{eq:recurNC}), we obtain the recursion relations
\begin{subequations}
\label{eq:recurC}
\begin{eqnarray}
t_{j}(1+i\alpha_{j,j+1}+i\beta_{j,j+1} e^{i\theta_{j,j+1}})+it_{j+1}\beta_{j,j+1} e^{i\theta_{j,j+1}}+t_{j-1}(i\alpha_{j,j+1}-1)=0 ,\  \textit{j \rm odd}, \ \\
t_{j}(1+i\alpha_{j-1,j})+t_{j-1}(i\alpha_{j-1,j}-1+i\beta_{j-1,j}e^{i\theta_{j-1,j}})+it_{j-2}\beta_{j-1,j}e^{i\theta_{j-1,j}}=0, \  \textit{j \rm even}, \ 
\end{eqnarray}
\end{subequations}
which can be solved to obtain the transmission and reflection coefficients.
Here
\begin{equation}
\alpha_{j,j+1}=\frac{\delta\omega\Gamma_{j}}{\delta\omega^2-J^{2}_{j,j+1}}, \quad \beta_{j,j+1}=\frac{J_{j,j+1}\sqrt{\Gamma_{j}\Gamma_{j+1}}}{\delta\omega^2 -J^{2}_{j,j+1}}.
\end{equation}
For the case of a single dimer the transmission coefficient is given by
\begin{equation}
t=\frac{4ie^{i\theta}J\Gamma+4e^{2i\theta}J^2+e^{2i\theta}(\gamma-\Gamma-2i\Delta)^{2}}{-4ie^{i\theta}J\Gamma+4J^2+(\gamma+\Gamma-2i\Delta)^{2}},
\end{equation}
where the detuning $\Delta=\omega-\Omega$.
We recall that the single two-level atom transmission coefficient is given by \cite{mirza2017chirality}
\begin{equation}
t_{A}=\Bigg(\frac{\gamma-\Gamma-2i\Delta}{\gamma+\Gamma-2i\Delta}\Bigg).
\end{equation}
Note that in the absence of interactions, when $J=0$, we see that $t=(e^{i\theta}t_{A})^{2}$, which corresponds to the cascaded transmission of two atoms.

\subsection{Periodic arrangement} 
Adopting the experimental scenario in \cite{cheng2017waveguide, akimov2007generation,*chang2006quantum}, we consider a Ag nanowire coupled to semiconductor quantum dots. The wavelength corresponding to the transition frequency of the quantum dot is $\lambda_{QD}=655$nm, the dimer length $L=32.75$nm and the optical wavelenth (corresponding to the surface plasmon mode) is $\lambda_{SP}=211.8$nm. We thus obtain $\theta=2\pi L/\lambda_{QD}=0.314$ and the dipole-dipole interaction is $J=46.2\Gamma_{0}$. We take the dimer separation to be $3L$, which implies that interaction between dimers $J_{d}\ll J$; it is therefore permissible to neglect the interaction between dimers.

\begin{figure}[t]\centering
  \begin{tabular}{@{}cccc@{}}
   \hspace{-2mm}\includegraphics[width=2.3in, height=1.6in]{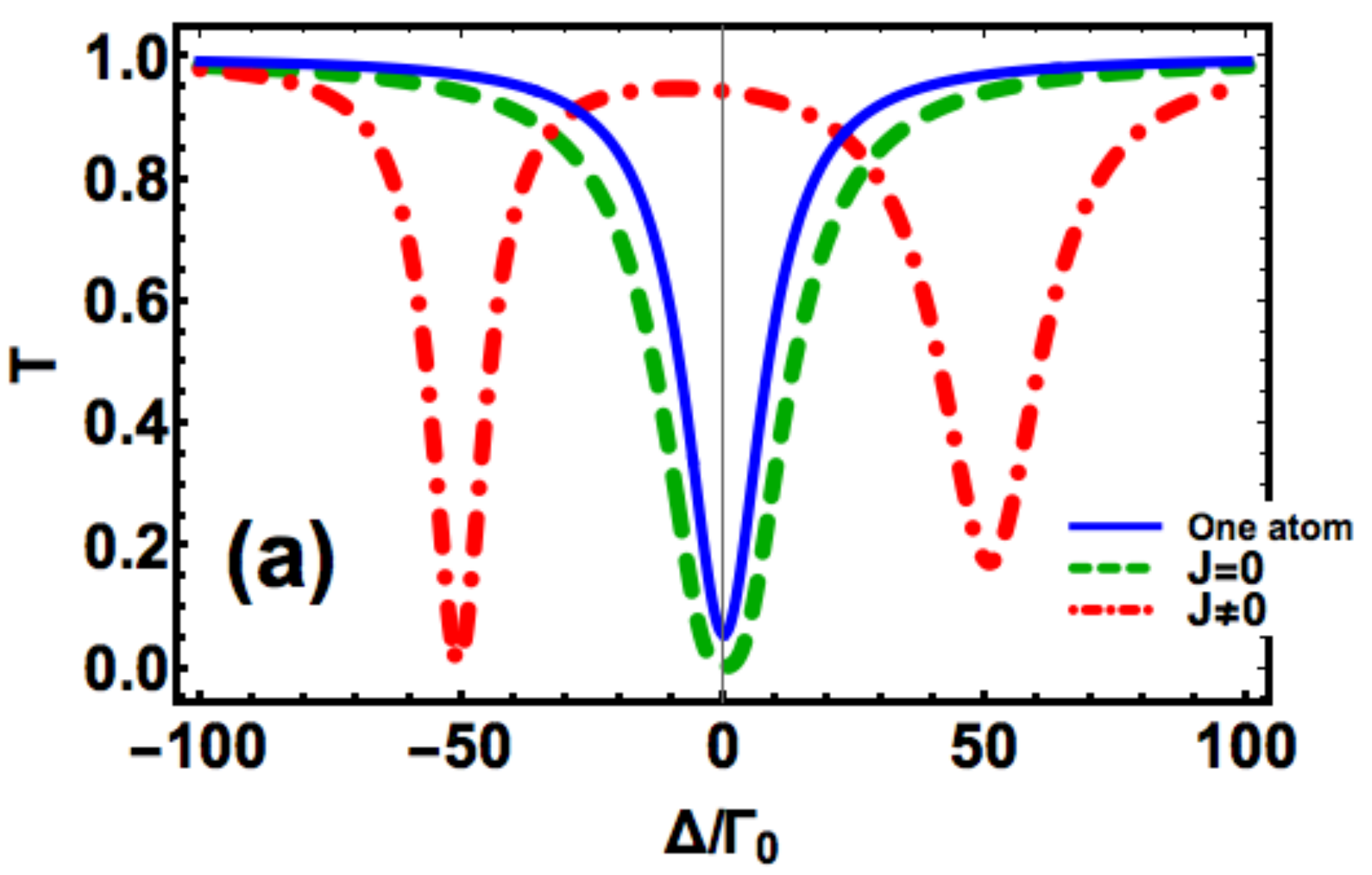} 
  \hspace{-1mm}\includegraphics[width=2.3in, height=1.6in]{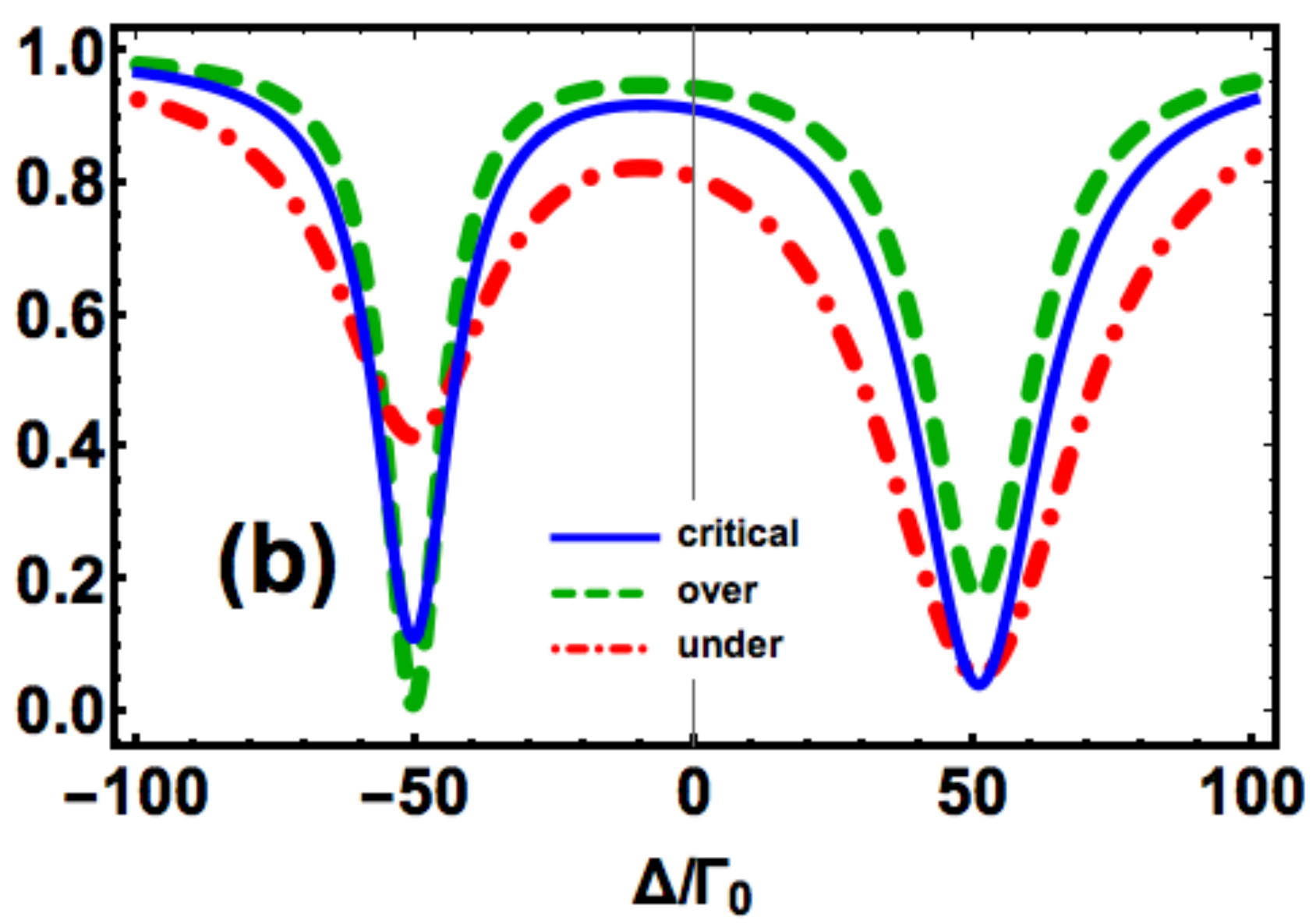}
   \hspace{-1mm}\includegraphics[width=2.3in, height=1.65in]{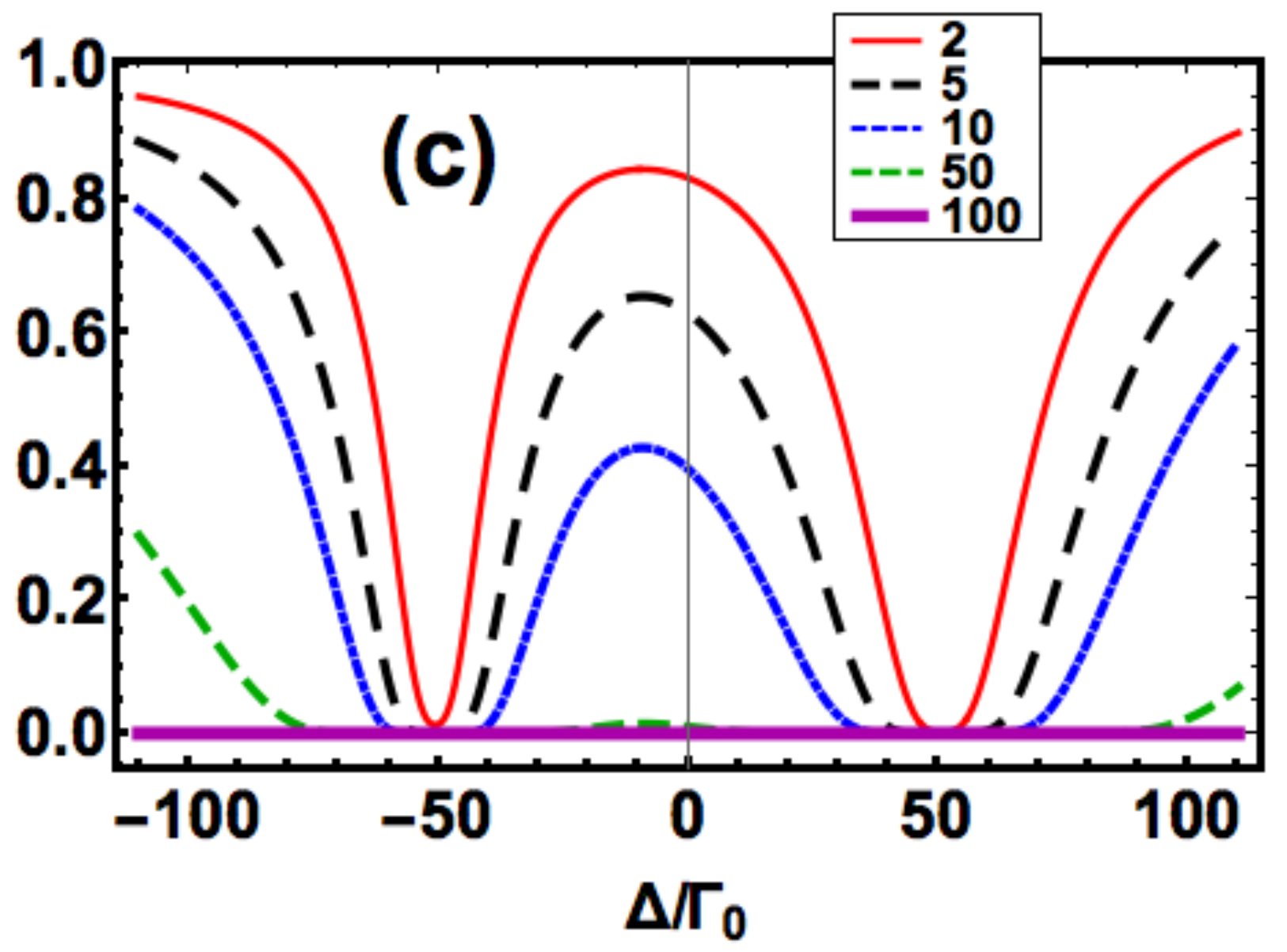}
  \end{tabular}
\captionsetup{
  format=plain,
  %margin=1em,
  justification=raggedright,
  singlelinecheck=false
}
\caption{(Color online) Transmission plots in the periodic chiral case. 
A single dimer is considered in (a) and (b). Multiple dimers are considered in (c). The parameters $\theta=0.314$ and $J=46.02\Gamma_{0}$ are the same in all plots. The other parameters are (a) $\gamma=6.86\Gamma_{0}$ and $\Gamma=11.103\Gamma_{0}$ (b) Over-coupled: $\gamma=6.86\Gamma_{0}$ and $\Gamma=11.103\Gamma_{0}$, Under-coupled: $\gamma=26.86\Gamma_{0}$ and $\Gamma=11.103\Gamma_{0}$, and critically-coupled: $\gamma=\Gamma=11.103\Gamma_{0}$ (c) $\gamma=\Gamma=11.103\Gamma_{0}$.}
\label{Fig2}
\end{figure}

In Fig.~\ref{Fig2} we examine the behavior of the transmission coefficient $T$ as a function of the detuning $\Delta$ and the number of dimers $n$.
In Fig.~\ref{Fig2}(a), we consider the case of a single dimer, where we find that in the absence of interactions, $T$ takes its smallest value at $\Delta=0$. The single atom transmission is also shown for comparison.
In the presence of interactions, the transmission curves develops two asymmetric peaks. In part Fig.~\ref{Fig2}(b) we plot the transmission in the three regimes:
undercoupled ($\gamma>\Gamma$), overcoupled ($\gamma <\Gamma$) and critically coupled ($\gamma=\Gamma$). We find that unlike the single atom case (where the transmission takes the smallest value at resonance in the critically-coupled regime)~\cite{mirza2017chirality}, for the dimer problem, only the resonance with positive detuning takes the lowest value in the critically-coupled regime. Finally, in part Fig.~\ref{Fig2}(c) we plot the transmission for a multi-dimer chain in the critically-coupled regime. We observe that as the number of dimers increases, the width of the resonances increases. Eventually for 100 dimers, the peak widths grow to such an extent that the separation between the peaks vanishes, and a wide region of null transmission appears.

\begin{figure}[t]\centering
  \begin{tabular}{@{}cccc@{}}
   \hspace{-2mm}\includegraphics[width=2.75in, height=1.9in]{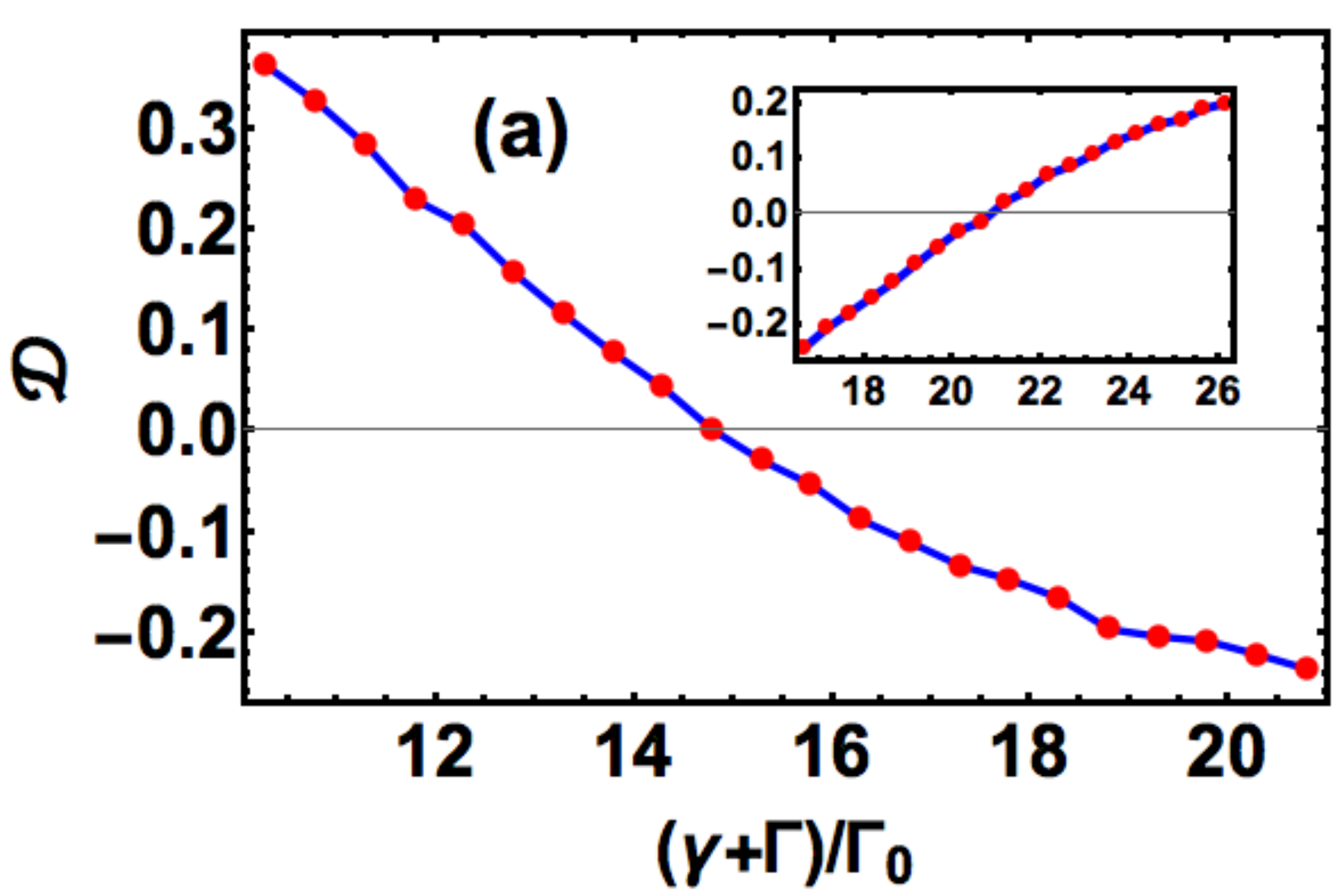} 
  \hspace{2mm}\includegraphics[width=2.75in, height=1.9in]{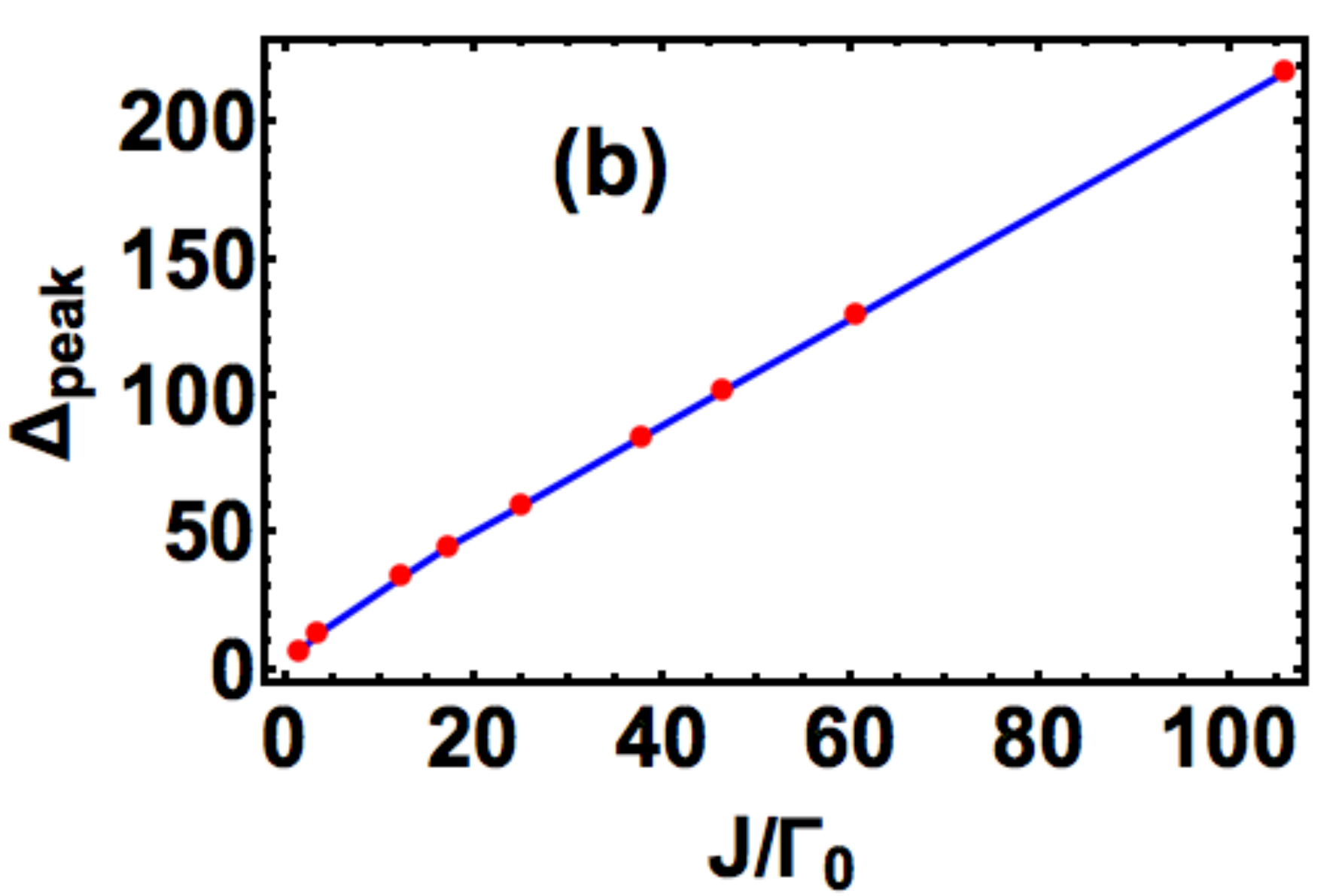}
  \end{tabular}
\captionsetup{
  format=plain,
  margin=1em,
  justification=raggedright,
  singlelinecheck=false
}
\caption{(Color online) (a) Dependence of the difference in peak heights $\mathcal{D}$ on $\gamma+\Gamma$. (b) Effect of $J$ on peak separation $\Delta_{\rm peak}$. Except for $\gamma$ and $\Gamma$ values in (a) and $J$ value in (b), all other parameters are the same as used in Fig.\ref{Fig2}.}
\label{Fig3}
\end{figure}

Next we consider the effect of varying the coupling parameter $J$ on the separation of the transmission peaks. When either $\gamma=0$ or $\Gamma=0$, then $T=1$ independent of $\Delta$. However, if we increase the parameter $\gamma+\Gamma$ then for certain choices of $\gamma$ and $\Gamma$ the peak in transmission becomes symmetric. To this end, in Fig.\ref{Fig3}(a) we plot the difference in peak heights $\mathcal{D}$ as we vary $\gamma+\Gamma$. We set $\gamma=6.86\Gamma_{0}$ and vary $\Gamma$ from $\gamma/2$ to almost $2\gamma$. In the inset, we put $\Gamma=11.103\Gamma_{0}$ and vary $\gamma$ from $\Gamma/2$ to $2\Gamma$. In both curves, we observe that $\mathcal{D}=0$ can be achieved. In Fig.\ref{Fig3}(b) we plot the peak separation $\Delta_{\rm peak}$ as a function of dipole-dipole interaction. We observe that the peak separation increases as the value of $J$ is increased.  

\subsection{Effects of disorder}
We now consider the effects of disorder in multidimer chains.  Our goal is to identify conditions which give rise to photon localization. In what follows, all random variables are assumed to be Gaussian distributed with a probability density of the form
\begin{equation}
\label{gaussian}
P(x)=\frac{1}{\sqrt{2\pi\sigma^2}}e^{-(x-\overline{x})^{2}/2\sigma^{2}},
\end{equation}
where  $\overline{x}$ is the mean and $\sigma$ is the standard deviation of the random variable $x$. We will sometimes refer to $\sigma$ as the strength of the disorder.  We study two models of disorder: random dimer lengths and random dimer separations.

\begin{figure}[t]
\centering
\includegraphics[width=2.85in, height=1.85in]{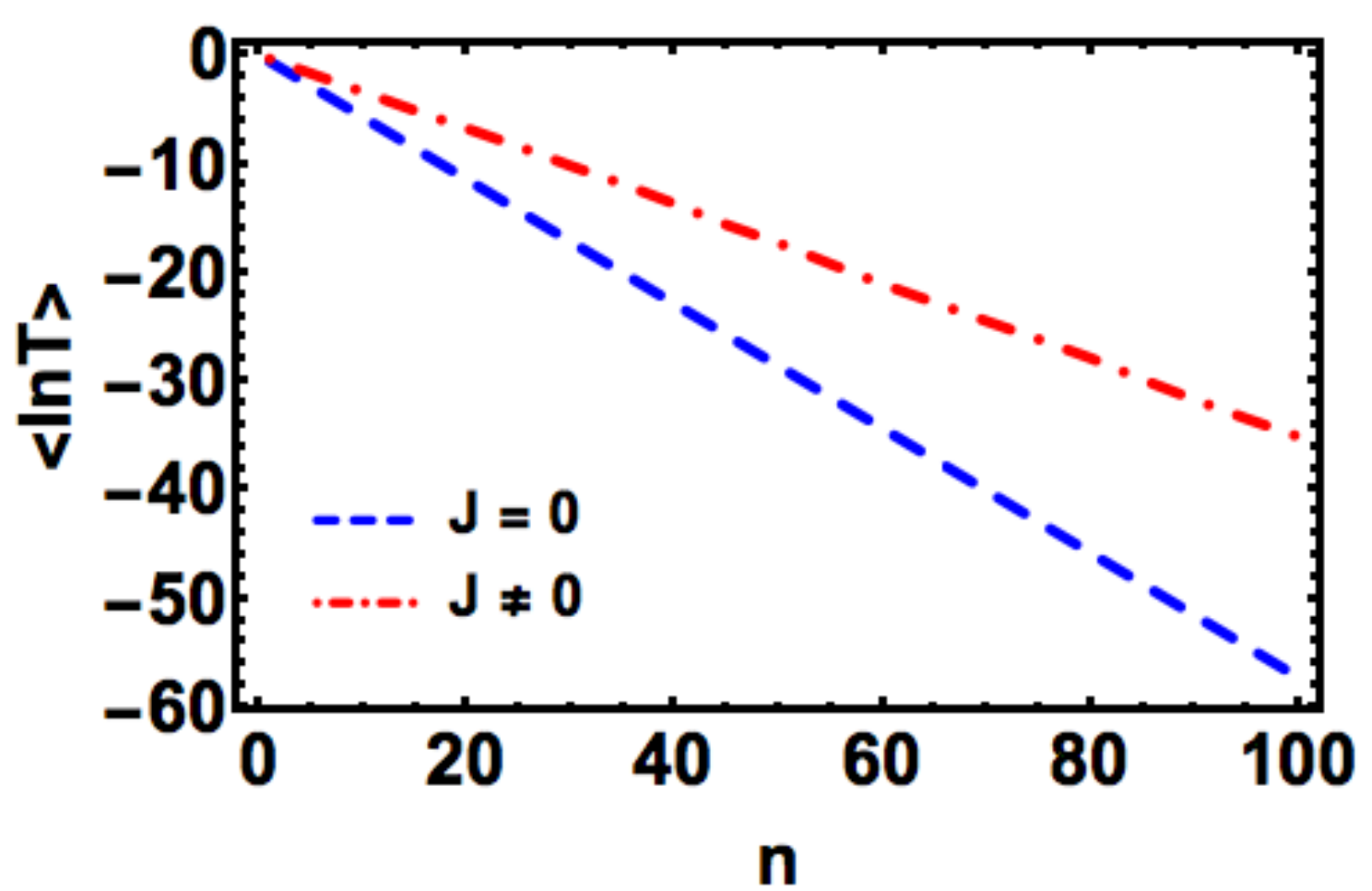} 
\captionsetup{
  format=plain,
  margin=1em,
  justification=raggedright,
  singlelinecheck=false
}
\caption{(Color online) Dependence of $\langle \ln T \rangle$ on the number of dimers $n$ for the case of random dimer lengths. 
We take $\gamma=6.86\Gamma_{0}$, $\Gamma=11.1033\Gamma_{0}$, $J=46.02\Gamma_{0}$, $\Delta=15\Gamma_{0}$, and $\sigma=0.2$. An average over $10^{5}$ realizations of the disorder has been performed. The error bars are too small to be displayed.}
\label{Fig4}
\end{figure}

\subsubsection{Localization}

Following standard procedures, we consider $\langle \ln T \rangle$ as a measure of photon transport in disordered systems~\cite{markos2008wave}.
Here the statistical average $\langle\cdots\rangle$ is carried out over
dimer lengths or separations, regarded as independent and identically distributed random variables. The corresponding localization length $\xi$ \cite{markos2008wave,izrailev1999localization} is defined as
\begin{equation}
\label{eq:llc}
\xi^{-1} = - \lim_{n\to\infty} \frac{\langle \ln T \rangle}{n},
\end{equation} 
where $n$ is the number of dimers.
We regard a numerical demonstration of the linear dependence of $\langle \ln T \rangle$ on $n$ as providing evidence for localization. A proof of localization in chiral waveguide QED is possible in some physical settings~\cite{mirza2017chirality,mirza2017influence}, but is in general an open problem. 

The dependence of $\langle \ln T\rangle$ on the number of dimers $n$ is shown in Fig.~\ref{Fig4} for the case of random dimer lengths. We see that 
the localization length is \emph{larger} in the presence of dipole-dipole interactions. In subsection~IV.B.2, we show this observation also extends to bidirectional waveguides when there is dimer separation disorder (see Fig.~\ref{Fig8}(d)).

\subsubsection{Disorder in dimer separation}
Suppose that the dimer length is fixed and that the separation between dimers is random. It follows from Eq.~(\ref{eq:recurC}) that the transmission $T$ does not dependent on the dimer separation. We conclude that the transmission of a chiral waveguide is \emph{immune} to disorder in dimer separation.  This result is consistent with our previous studies on single photon transport in chiral waveguides coupled to two-level atoms \cite{mirza2017chirality}. 

\subsubsection{Disorder in dimer length}
\begin{figure}[t]
\centering
  \begin{tabular}{@{}cccc@{}}
   \hspace{-2mm}\includegraphics[width=2.25in, height=1.75in]{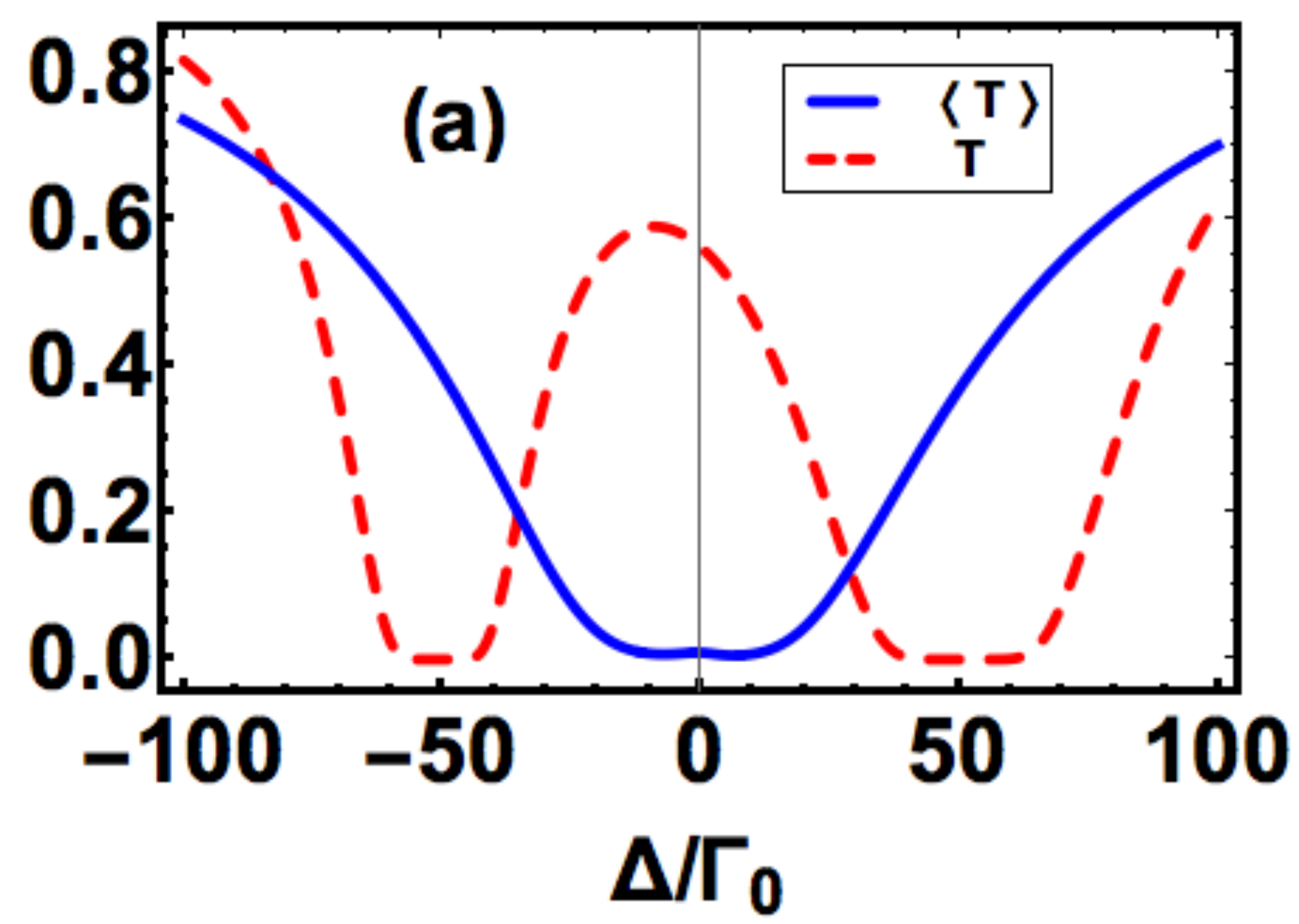} 
  \hspace{-2mm}\includegraphics[width=2.25in, height=1.75in]{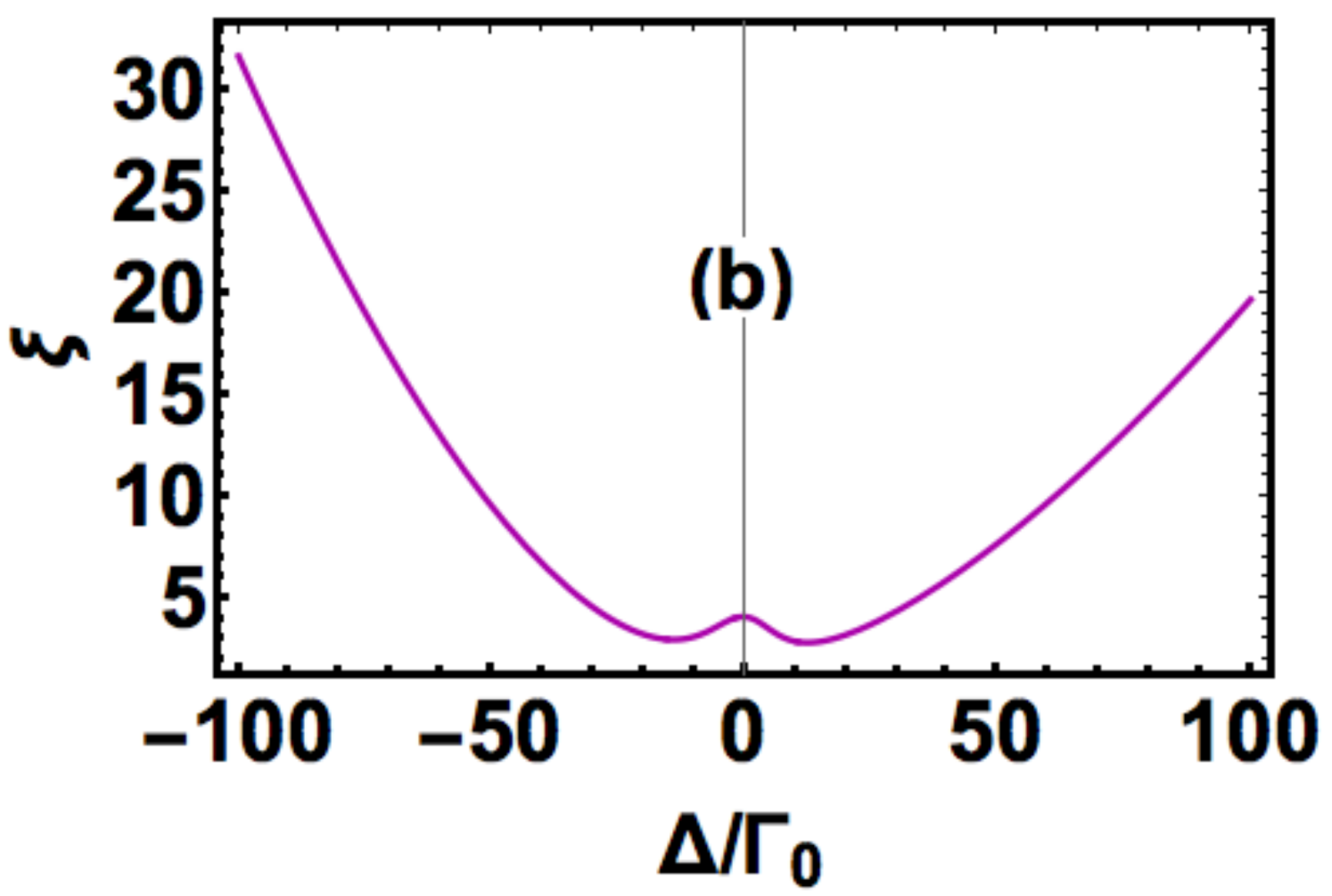}
   \hspace{-1mm}\includegraphics[width=2.3in, height=1.725in]{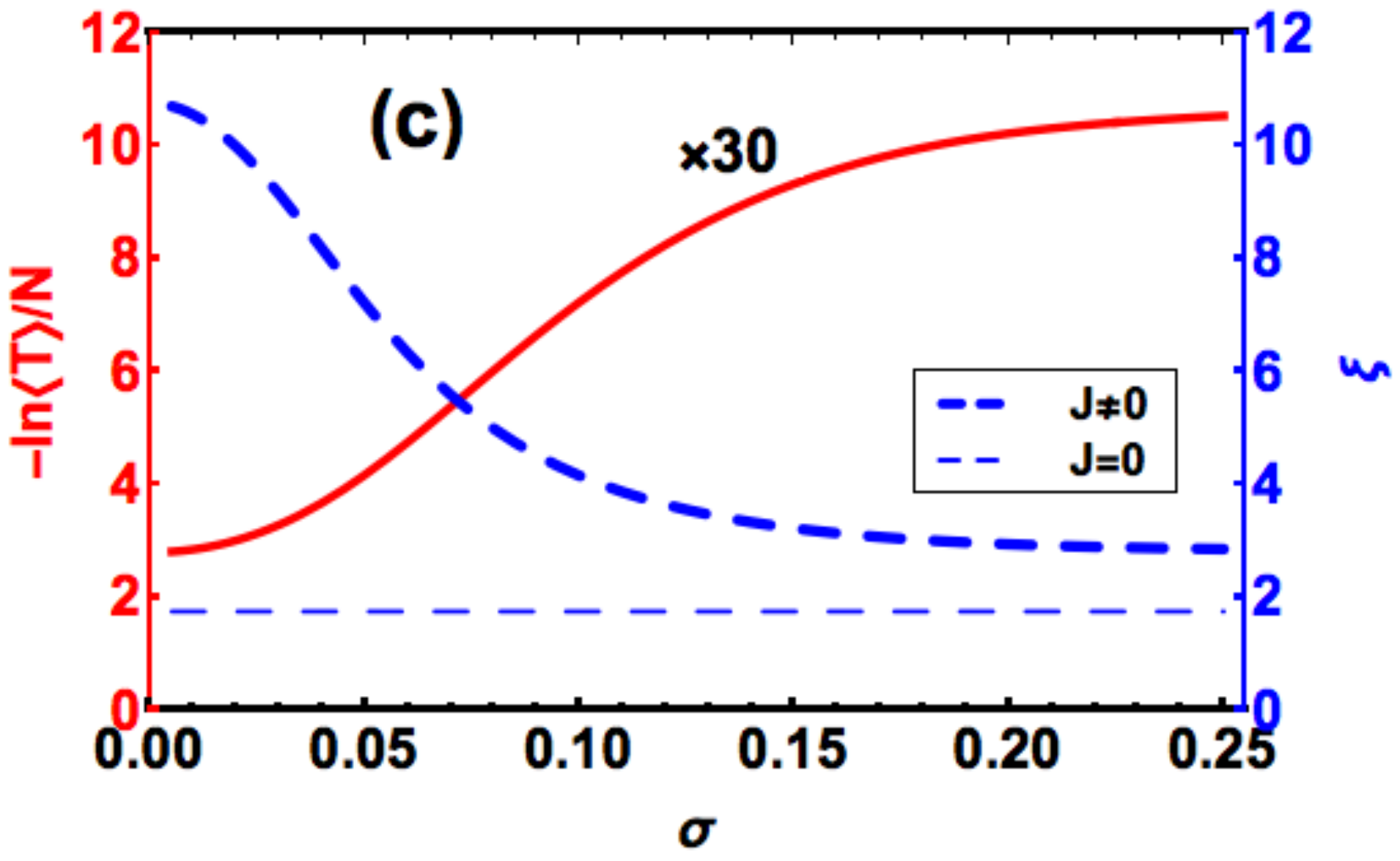}
  \end{tabular}
\captionsetup{
  format=plain,
  margin=1em,
 justification=raggedright,
  singlelinecheck=false
}
\caption{{(Color online) Influence of dimer-length disorder on single-photon transmission in a chiral waveguide. For both (a) and (b) we used the following parameters: mean dimer length $L=32.75$nm, fixed dimer separation $3L$, $\gamma=6.86\Gamma_{0}$ and $\Gamma=11.103\Gamma_{0}$.  In addition, $\sigma=0.25\times 2\pi L/\lambda_{QD}$. (a) Dependence of transmission on detuning in periodic and disordered chaings. A ten dimer chain is considered. (b) Dependence of localization length on detuning. (c) Dependence of localization length and average transmission on disorder strength $\sigma$, in units of $2\pi L/\lambda_{QD}$ with $\Delta=15\Gamma_{0}$; all other parameters are the same as used in (a).}}
\label{Fig5}
\end{figure} 

We now suppose that the dimer lengths are random. Here we assume that the dimer lengths $L_{j}$, defined by $L_{j}=x_{j}-x_{j-1}$,  are independent and identically distributed random variables. The net transmission from a chain of $n$ dimers is given by
\begin{eqnarray}
\langle T \rangle &=& \int \prod_{j=1}^n dL_{j} P(L_{j}) |T_j|^2 \\
\label{eq:TotalTC}&=& \langle |\tau|^2 \rangle^n.
\label{eq:avgT}
\end{eqnarray}
Here
\begin{equation}
\langle |\tau|^2 \rangle = \int d\delta P(L) |\tau|^2  ,
\end{equation}
where 
\begin{equation}
\label{eq:Tsd}
\tau = \frac{4ie^{ikL}J\Gamma+e^{2ikL}(4J^{2}+(\gamma-\Gamma-2i\Delta)^{2})}{(\gamma+\Gamma-2i\Delta)^{2}+4J(J-i\Gamma e^{ikL})} ,
\end{equation}
with $k\equiv 2\pi/\lambda_{QD}$. We note that the dimer length appears both in the dipole-dipole interaction $J$ as well as in the phase factors.
The average is performed over all dimer lengths. It is easy to see that
\begin{equation}
\langle \ln T \rangle = n \langle \ln |\tau|^2 \rangle ,
\end{equation}
and hence from Eq.~(\ref{eq:llc}) we find
\begin{equation}
\label{eq:lle}
\xi^{-1} = - \langle \ln |\tau|^2 \rangle .
\end{equation}
Using (\ref{eq:TotalTC}) and (\ref{eq:lle}) we can calculate the average transmission and localization length. In Fig.~\ref{Fig5}(a), we plot the transmission of a single photon in a disordered ten dimer chain (blue curve) with weak disorder. Note that strong disorder ($\sigma > 2\pi L/\lambda_{QD}$) leads to the presence of dimer-dimer interactions. For comparison, we have included the transmission for the corresponding periodic case (red dashed curve). We see that the presence of disorder considerably alters the transmission. The frequency doublet that arises in the periodic case disappears and a single transmission curve with a broad band of very small transmission is formed.
In Fig.~\ref{Fig5}(b), we plot the corresponding frequency-dependent localization length $\xi$. We observe that the localization length is smallest at the two frequencies near resonance. However, away from these points $\xi$ increases.

In Fig.~\ref{Fig5}(c) we plot the localization length and average transmission as a function of the disorder strength $\sigma$. We vary $\sigma$ between $0$ to $\pi L/2\lambda_{QD}$, so that the interaction between the dimers is negligible. We see that in the absence of disorder $T\simeq0.9$, which is consistent with Fig.~\ref{Fig2}(a). As expected, when $J=0$ (thin dashed curve), the localization length does not depend on $\sigma$. However, when $J \neq 0$ (thick dashed curve) an additional channel for photon transport opens up.

\section{Bidirectional waveguides}
We now consider the transmission characteristics of symmetric bidirectional waveguides. 
For a single dimer, the transmission ($t$) and reflection ($r$) coefficients take the form
\begin{subequations}
\label{eq:TIAENC}
\begin{eqnarray}
t=\frac{-4ie^{i\theta}J\Gamma+4ie^{3i\theta}J\Gamma-e^{2i\theta}(4J^{2}+(\gamma-2i\Delta)^{2})}{4ie^{i\theta}J\Gamma+4e^{2i\theta}\Gamma^{2}-(\gamma+2\Gamma-2i\Delta)^{2}-4J(J-i\Gamma e^{i\theta})    },\label{eq:phic}\\
r=\frac{2\Gamma(-4ie^{i\theta}J+\gamma+2\Gamma-e^{2i\theta}(-\gamma+2\Gamma+2i\Delta)-2i\Delta)}{4ie^{i\theta}J\Gamma +4e^{2i\theta}\Gamma^{2}-(\gamma+2\Gamma-2i\Delta)^{2}-4J(J-i\Gamma e^{i\theta}) }.
\end{eqnarray}
\end{subequations}

To explicitly demonstrate  the waveguide mediated coupling between atoms, we set $J=0$ and obtain 
\begin{subequations}
\label{eq:rtJzero}
\begin{eqnarray}
t &=& \frac{-e^{2i\theta}(\gamma-2i\Delta)^{2}}{4e^{2i\theta}\Gamma^{2}-(\gamma+2\Gamma-2i\Delta)^{2}   },\\
r &=& \frac{2\Gamma(+\gamma+2\Gamma-e^{2i\theta}(-\gamma+2\Gamma+2i\Delta)-2i\Delta)}{4e^{2i\theta}\Gamma^{2}-(\gamma+2\Gamma-2i\Delta)^{2} }.
\end{eqnarray}
\end{subequations}
The term proportional to $\Gamma^{2}$ in the denominator of (\ref{eq:rtJzero}) accounts for the waveguide mediated interaction between atoms, and is present even when $J=0$.

\subsection{Periodic arrangement}
\begin{figure}[t]
\centering
  \begin{tabular}{@{}cccc@{}}
   \hspace{-1mm}\includegraphics[width=2.25in, height=1.75in]{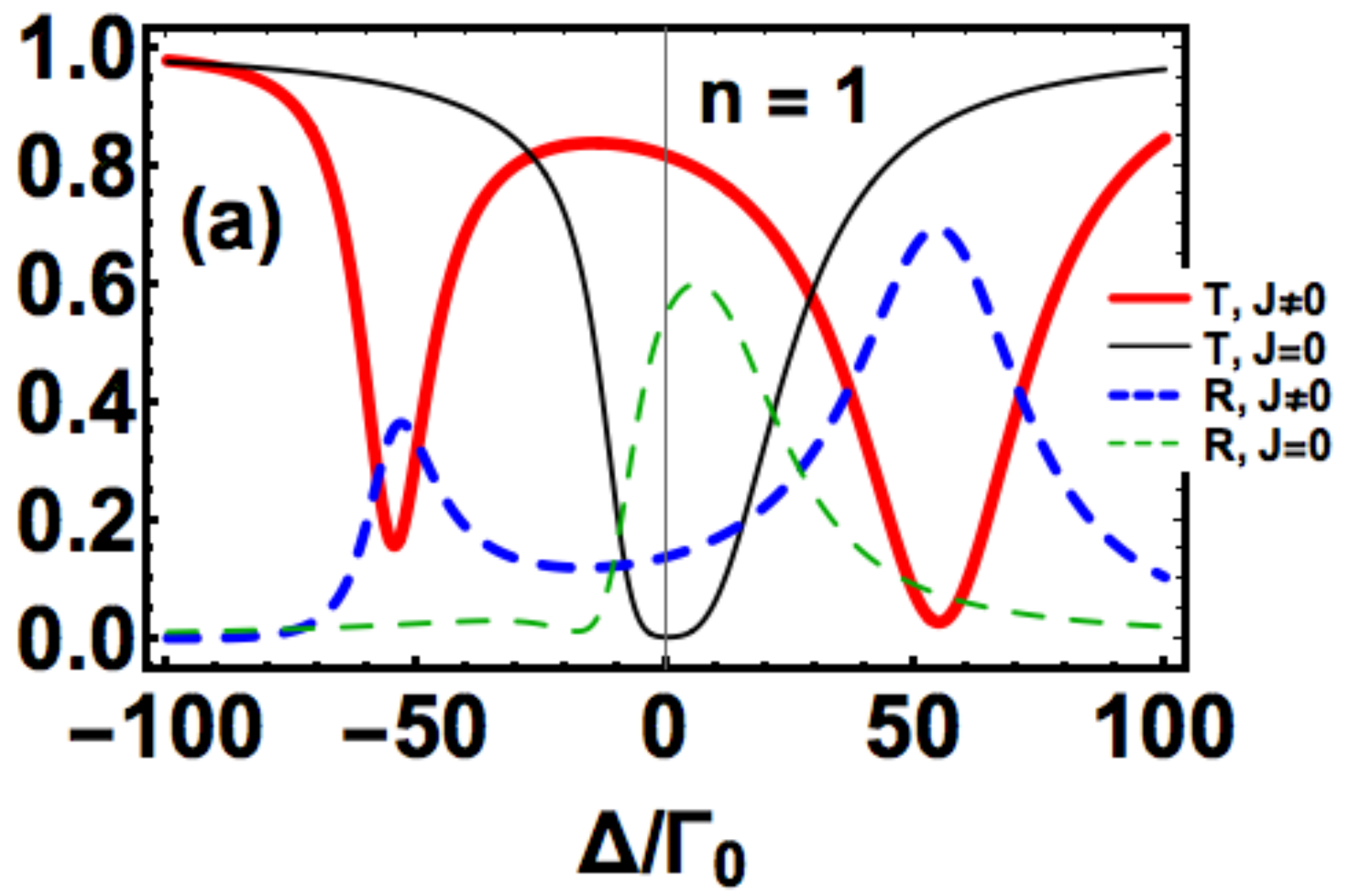} 
  \hspace{-1mm}\includegraphics[width=2.25in, height=1.75in]{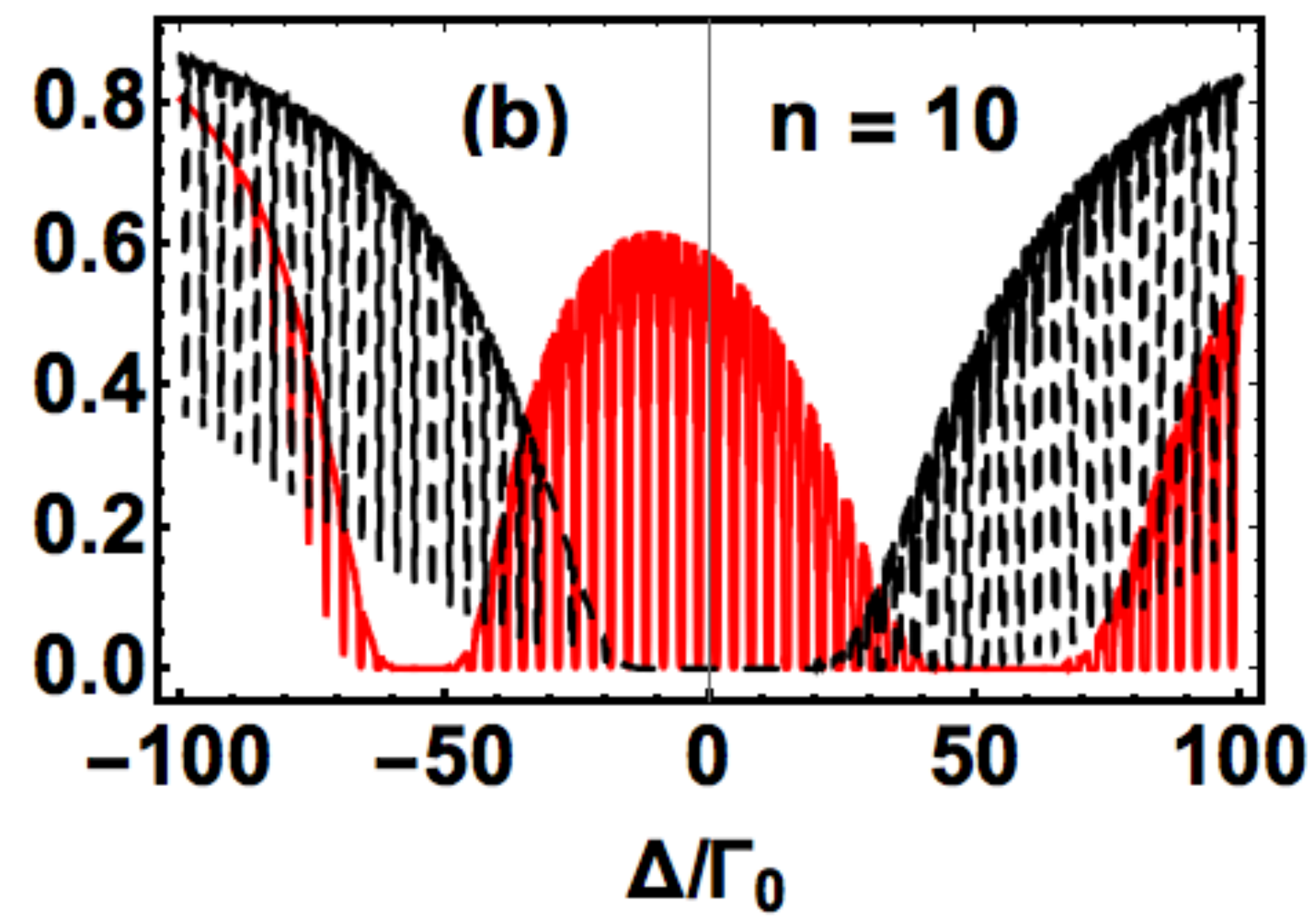}
   \hspace{-1mm}\includegraphics[width=2.25in, height=1.75in]{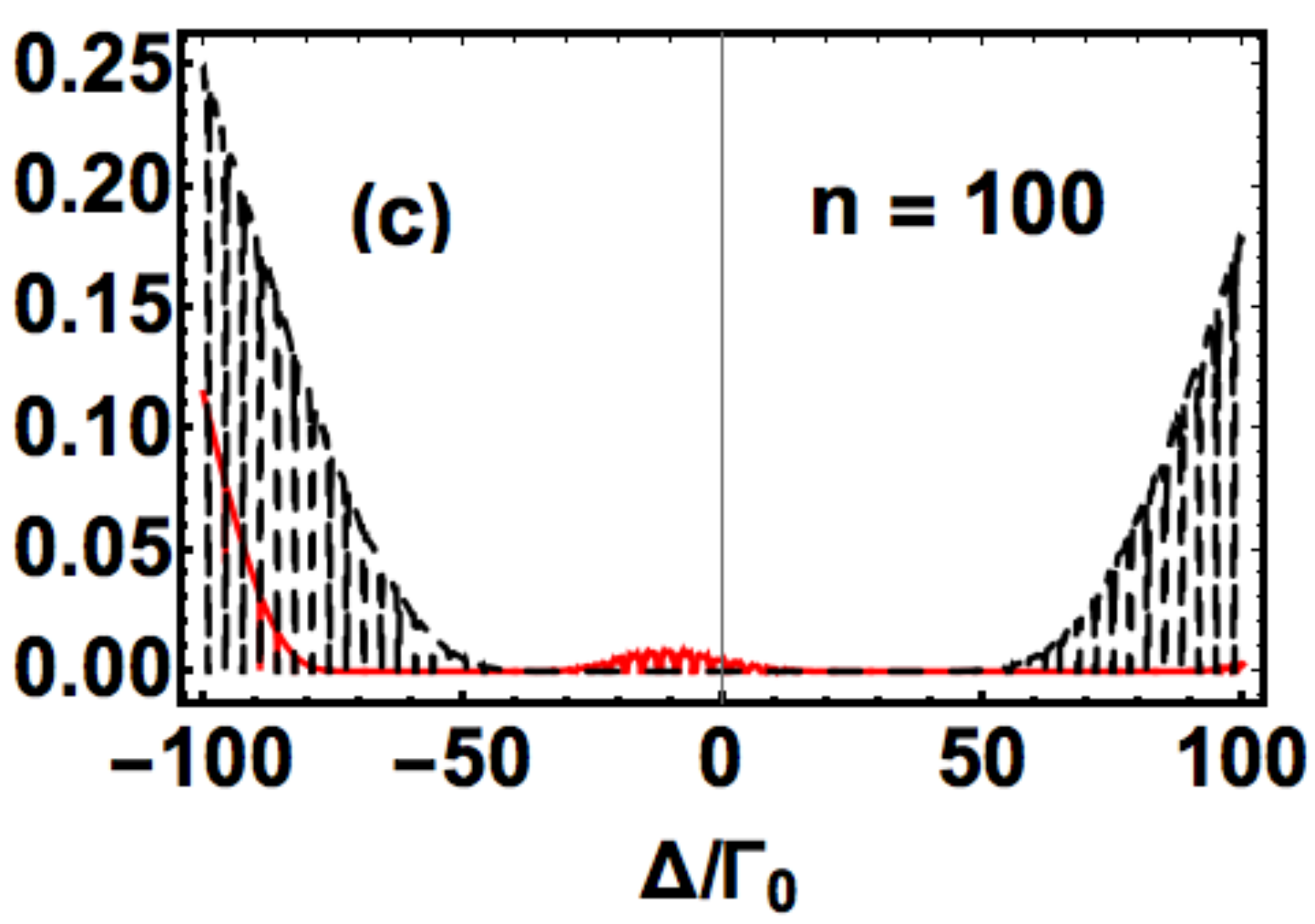}
  \end{tabular}
\captionsetup{
  format=plain,
  margin=1em,
  justification=raggedright,
  singlelinecheck=false
}
\caption{(Color online) Transmission plots in the periodic bidirectional case with and without dipole-dipole interaction. A single dimer is considered in (a), 10 dimers in (b), and 100 dimers in (c). Parameters in all plots: $\gamma=6.86\Gamma_{0}$, $\Gamma=11.103\Gamma_{0}$, $J=46.2\Gamma_{0}, L=32.75$nm and dimer separation $3L$.}
\label{Fig6}
\end{figure}
In Fig.\ref{Fig6}(a) we plot the transmission and reflection coefficients for a single dimer when $J=0$ and when $J\neq 0$. In the former case, the transmission takes a minimum value at $\Delta=0$ but exhibits a asymmetric Fano-like spectrum~\cite{cheng2012fano}. In the latter case, the spectrum is a frequency doublet with two asymmetric peaks. Similar to the single dimer chiral case, the splitting is due to dipole-dipole interactions. The loss due to non-zero $\gamma$ breaks the symmetry in peak heights. In particular, when $\gamma=0$ the peaks reside at $\Delta=\pm\sqrt{2\Gamma J\sin\theta+J^2}$~\cite{cheng2017waveguide}. In Fig.\ref{Fig6}(b) and Fig.\ref{Fig6}(c), we consider the setting of multdimer chains. For comparison, we also plot the transmission for $J=0$. We note that for $J\neq0$, in contrast to the chiral case, multiple narrow resonances emerge as the number of dimers is increased. We observe that the period of the narrow resonances is determined by the phase $\theta_{j,j+1}$ associated with the dimer length. An envelope appears in two regions centered at $\sim -60\Gamma_{0}$ and $\sim 60\Gamma_{0}$ with zero transmission. In contrast, for $J=0$ the zero transmission region is centered around the resonance and narrow resonances emerge away from $\Delta=0$. As the number of dimers increases, the band centered around $\Delta=0$ shows an order of magnitude reduction in amplitude for $J\neq0$.

\begin{figure}[t]
\centering
  \begin{tabular}{@{}cccc@{}}
\includegraphics[width=2.75in, height=1.85in]{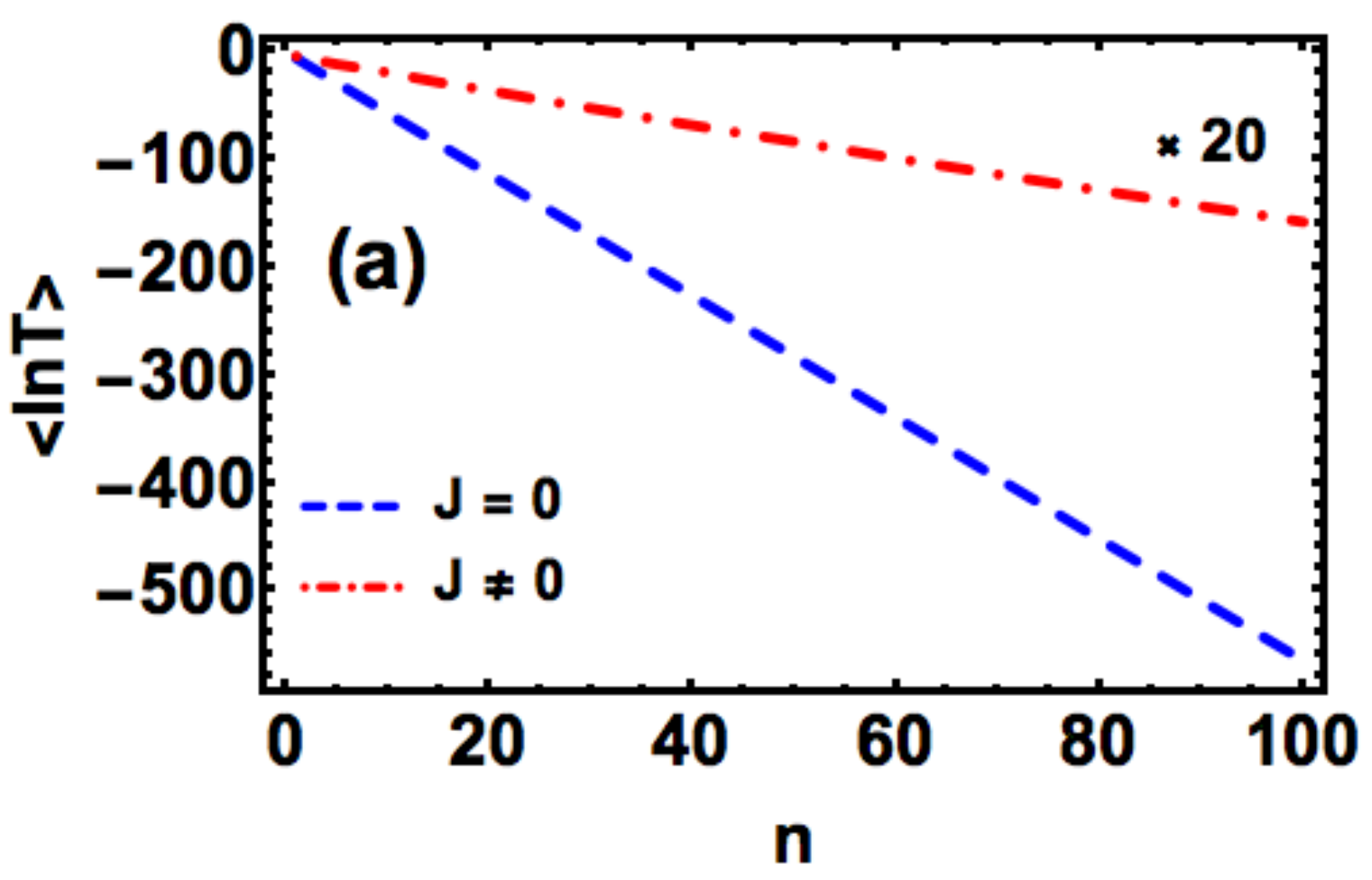} 
\hspace{2mm}\includegraphics[width=2.75in, height=1.85in]{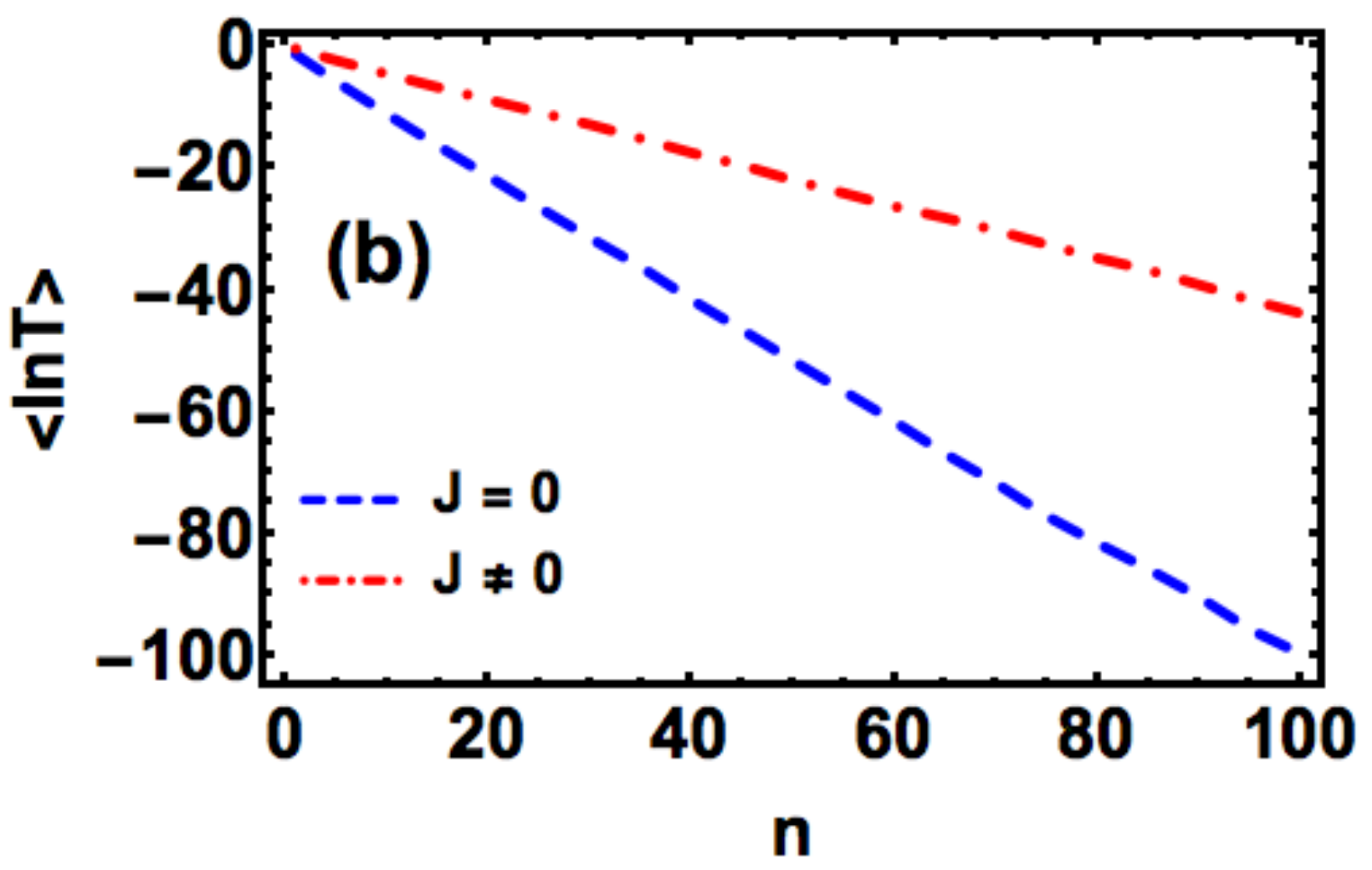} 
  \end{tabular}
\captionsetup{
  format=plain,
  margin=1em,
  justification=raggedright,
  singlelinecheck=false
}
\caption{(Color online) Dependence of $\langle \ln T \rangle$ on the number of dimers $n$. The case of disordered dimer separation is shown in (a) and disordered dimer lengths is shown in (b).
The following parameters were employed:  $\Gamma_{L}=\Gamma_{R}=11.103\Gamma_{0}$, $\gamma=6.86\Gamma_{0}$, $J=46.02\Gamma_{0}$, $\Delta=15\Gamma_{0}$, mean dipole separation $3L$ and $\sigma=0.2$.  Averages were carried out over $10^{5}$ realizations of the disorder. The error bars are too small to be displayed.}
\label{Fig7}
\end{figure}

\subsection{Effects of disorder}
\subsubsection{Localization}
We seek numerical evidence for localization. Fig.~(\ref{Fig7}) presents plots of $\langle \ln T\rangle$ as a function of the number of dimers $n$. We consider two cases of interest: disordered dimer separations and disordered dimer lengths. In each case, we demonstrate the effect of dipole-dipole interactions.

\begin{figure}[t]
\centering
  \begin{tabular}{@{}cccc@{}}
   \includegraphics[width=2.5in, height=1.75in]{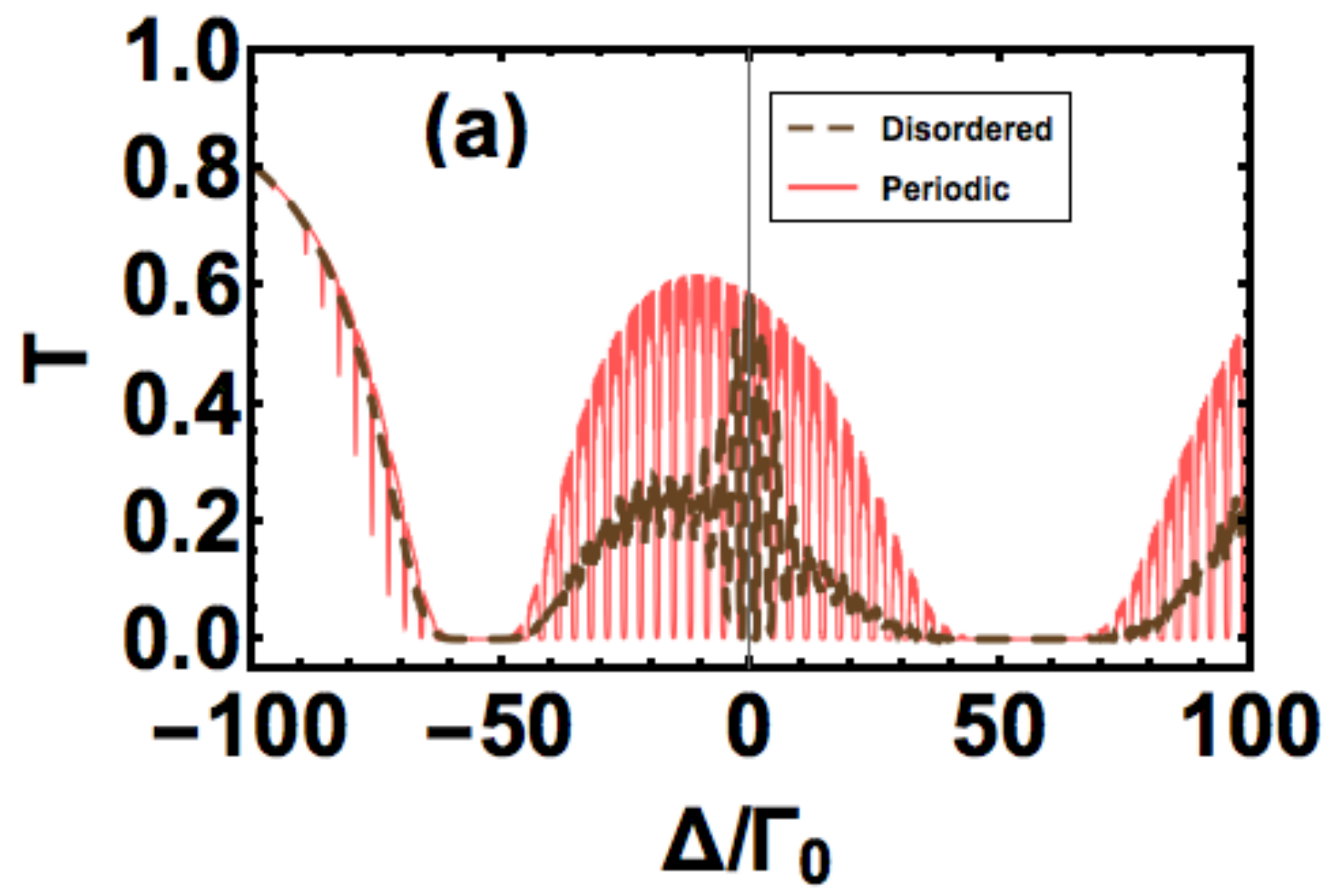} 
   \hspace{3mm}\includegraphics[width=2.5in, height=1.75in]{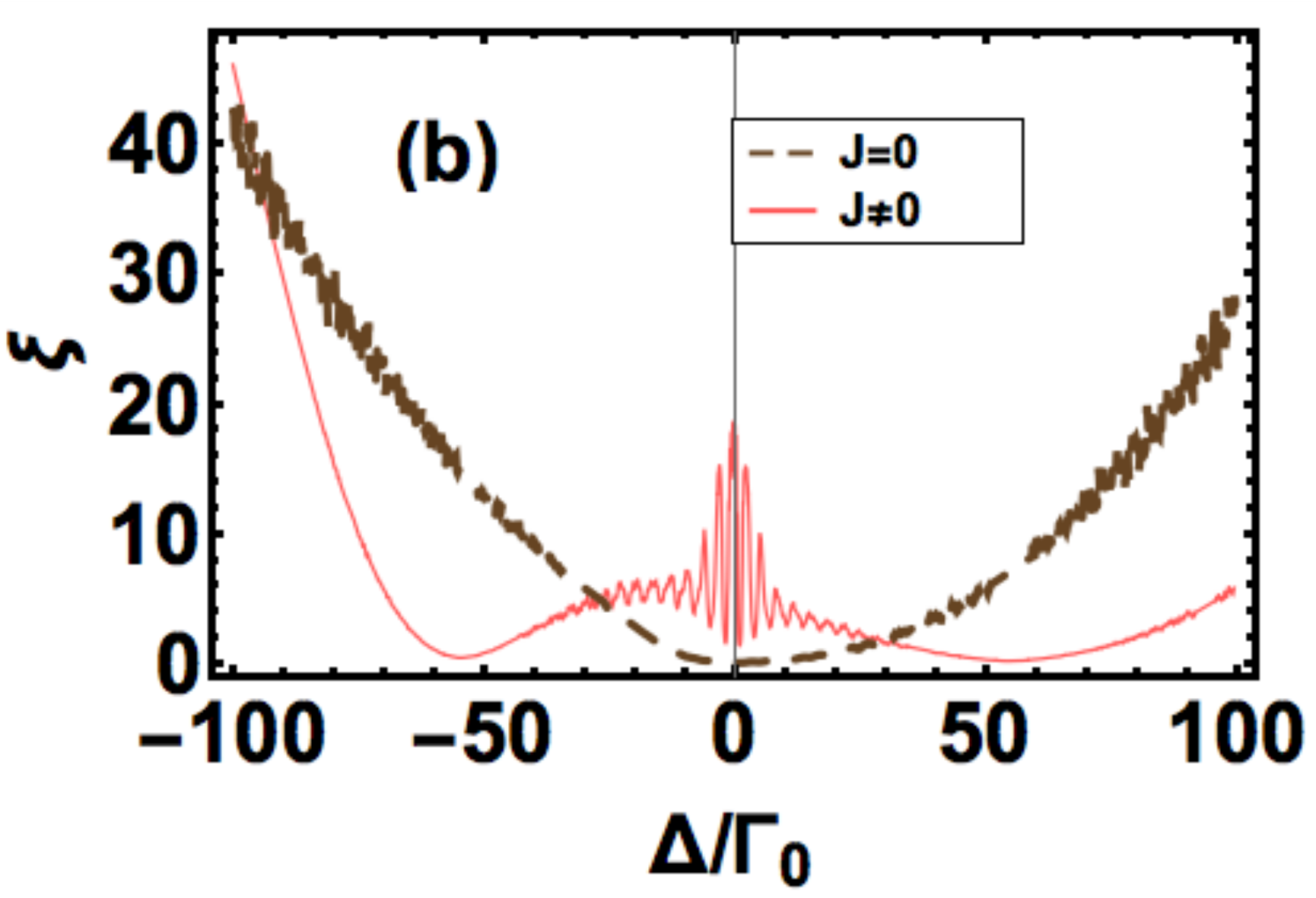}\\
     \includegraphics[width=2.5in, height=1.75in]{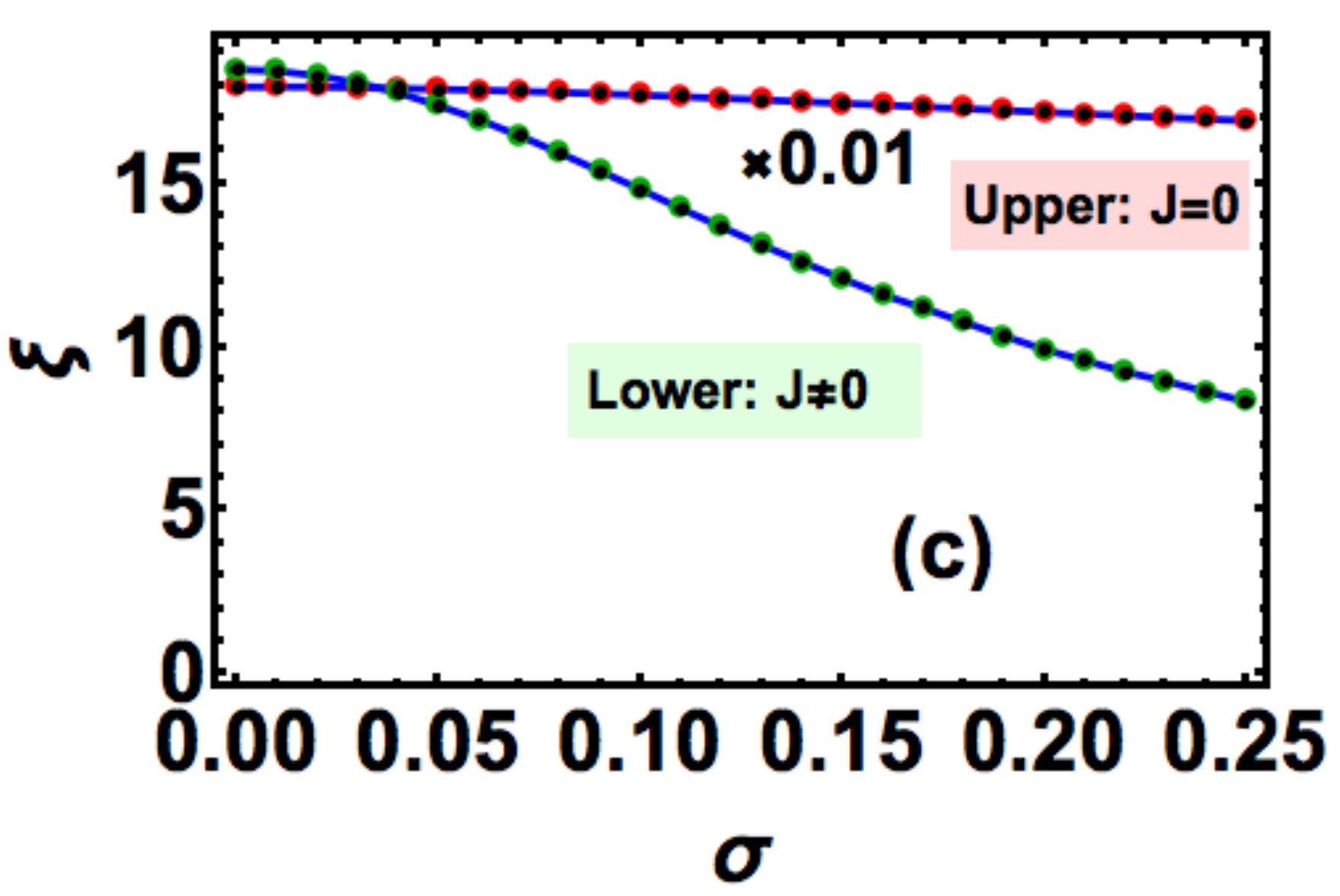}
     \hspace{3mm}\includegraphics[width=2.5in, height=1.75in]{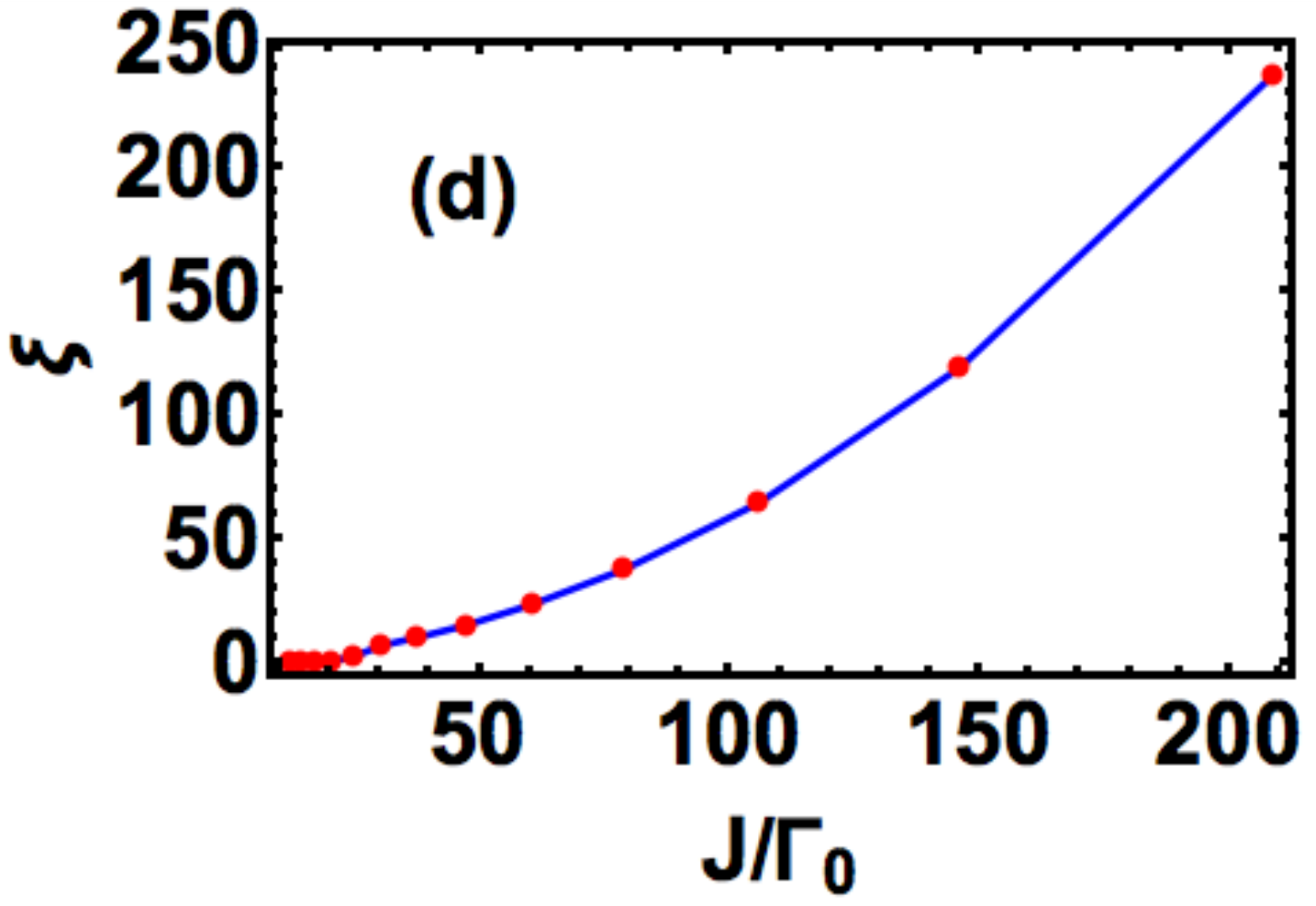}
  \end{tabular}
\captionsetup{
  format=plain,
  margin=1em,
  justification=raggedright,
  singlelinecheck=false
}
\caption{(Color online) Influence of dimer-separation disorder on single-photon transmission in a symmetric waveguide. (a) Dependence of transmission coefficient on detuning $\Delta$ in periodic and disordered systems. Parameters are $\gamma=6.86\Gamma_{0}$, $\Gamma=11.103\Gamma_{0}$, dimer length is $L=32.75$nm, and mean dimer separation is $3L$. (b) Dependence of localization length $\xi$ on detuning $\Delta$. The parameters are the same as in (a). In (a) and (b) we have chosen $\sigma=0.25 L$ with the average carried out over $500$ realizations of the disorder. (c) Dependence of $\xi$ on $\sigma$. (d) Dependence of $\xi$ on $J$. In both (c) and (d) we used the same parameters as in (a), and the average is performed over $10^{5}$ realizations with $n=100$. The error bars are too small to be displayed.}
\label{Fig8}
\end{figure}

\subsubsection{Disorder in dimer separation}
Suppose that the dimer length is fixed and that the separation between dimers is random. In Fig.~\ref{Fig8}(a) we plot the transmission coefficient as a function of detuning $\Delta$ for a ten dimer chain. For comparison, we also plot the transmission for a periodic chain. In Fig.~\ref{Fig8}(b) we plot the corresponding localization length $\xi$. We note that both the transmission coefficient and localization length show oscillatory behavior near resonance. When $J=0$, we find that the localization length takes its minimum value around $\Delta=0$, consistent with the case of position-disordered chains in bidirectional waveguides \cite{mirza2017chirality}. In Fig.~\ref{Fig8}(c) we exhibit the dependence of $\xi$ on the disorder strength $\sigma$. We find that $\xi$ decays with increasing $\sigma$ for both $J=0$ and $J\neq 0$. However, the scale of $\xi$ is nearly twice as large as for dimer-length disorder, as shown in Fig.~\ref{Fig9}(c). Finally, in Fig.~\ref{Fig8}(d) we plot the dependence of localization length on $J$. We select the values of $J$ corresponding to dimer lengths from $20$nm to $50$nm. As in the chiral case, we see that the localization length is larger in the presence of interactions.

\subsubsection{Disorder in dimer length}
\begin{figure}[t]
\centering
  \begin{tabular}{@{}cccc@{}}
   \hspace{-5mm}\includegraphics[width=2.25in, height=1.75in]{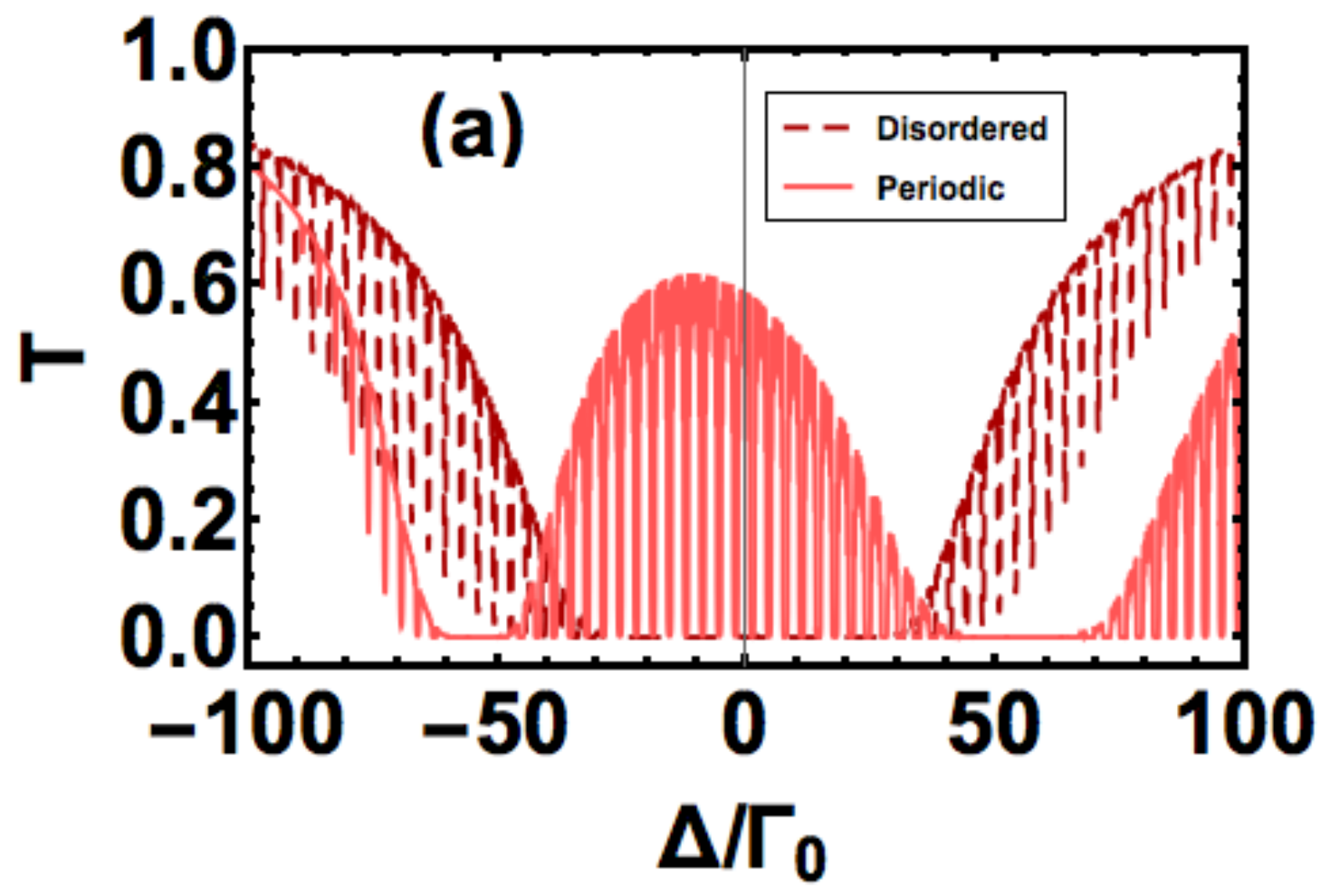} 
   \hspace{-2mm}\includegraphics[width=2.25in, height=1.75in]{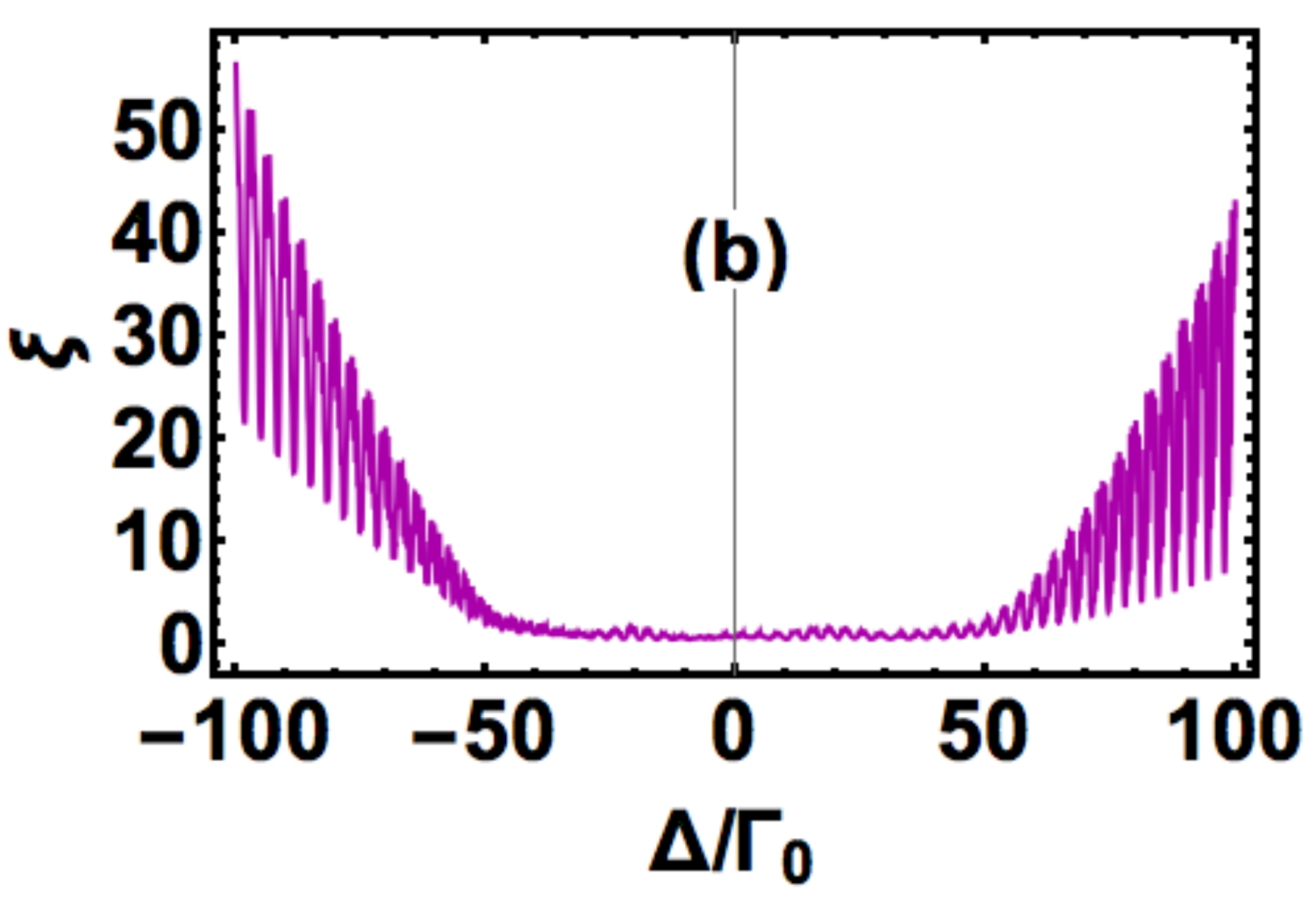}
   \hspace{-2mm}\includegraphics[width=2.25in, height=1.75in]{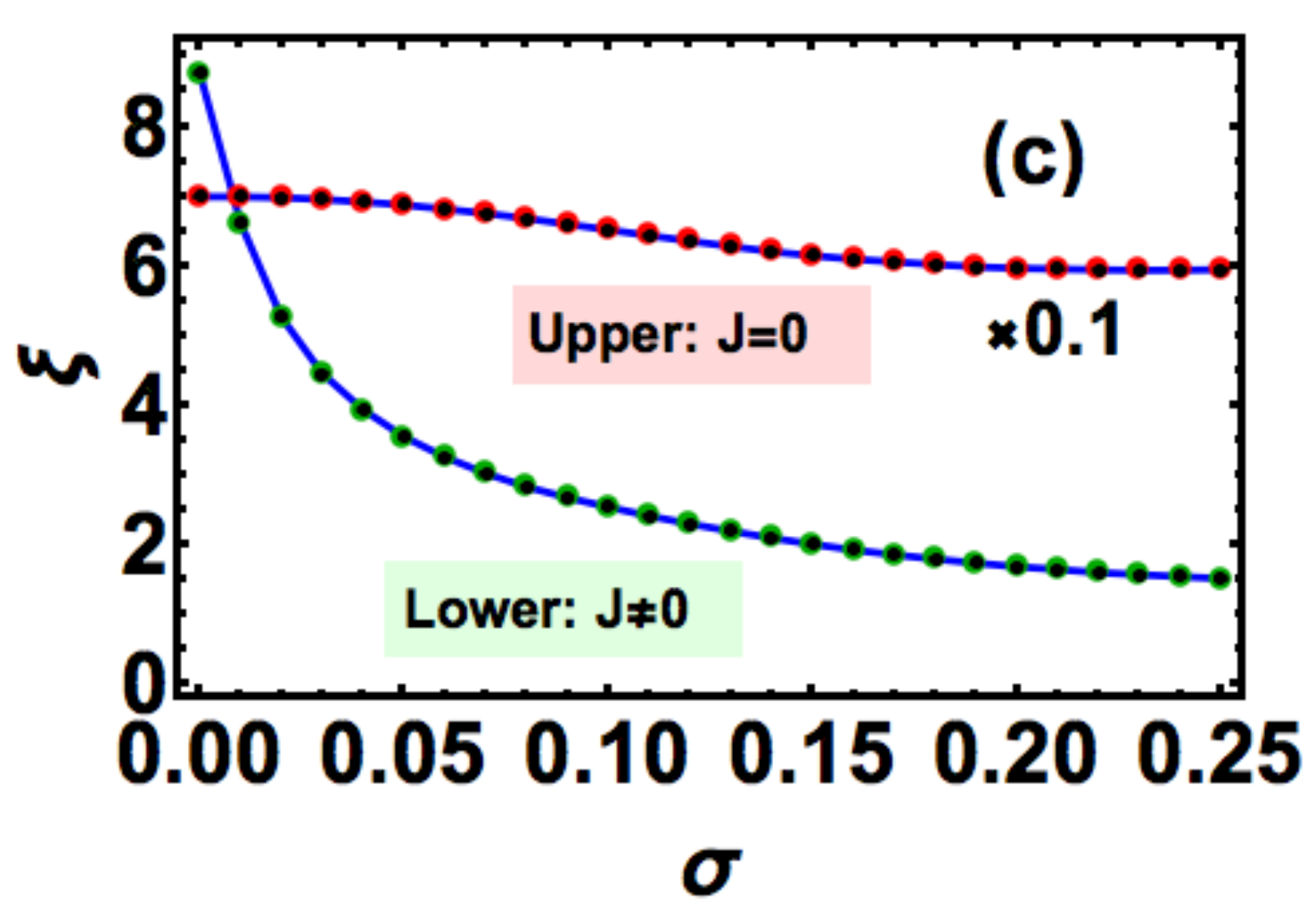}
  \end{tabular}
\captionsetup{
  format=plain,
  margin=1em,
  justification=raggedright,
  singlelinecheck=false
}
 \caption{(Color online) Influence of dimer-length disorder on single-photon transmission in a symmetric waveguide. (a) Dependence of transmission coefficient on detuning $\Delta$ in periodic and disordered systems. Parameters are $\gamma=6.86\Gamma_{0}$, $\Gamma=11.103\Gamma_{0}$, mean dimer length $L=32.75$nm, and fixed dimer separation $98.25$nm.  (b) Dependence of localization length $\xi$ on detuning. In (a) and (b) we have chosen $\sigma=0.25 L$ with the average carried out over $500$ realizations of the disorder. (c) Dependence of $\xi$ on $\sigma$ for the same parameters used in (a). The average over disorder is performed over $10^{5}$ realizations with $n=100$. The error bars are too small to be displayed.}\label{Fig9}
\end{figure}
We now suppose that the dimer lengths are random. In Fig.~\ref{Fig9}(a) we plot the transmission coefficient as a function of detuning $\Delta$ for a ten dimer chain. For comparison, we also plot the transmission for a periodic chain. We observe that the presence of disorder alters the profile of the transmission substantially. For instance, the two regions of suppressed transmission obtained in the periodic arrangement shift into a single band of perfect reflection centered around $\Delta=0$. The envelopes of several resonances around the null transmission survive even in the presence of disorder. In Fig.~\ref{Fig9}(b) we plot the corresponding localization length $\xi$. We note that such an oscillatory profile does not occur in the absence of dipole-dipole interactions~\cite{mirza2017chirality}. Finally, in Fig.~\ref{Fig9}(c) we exhibit the dependence of $\xi$ on the disorder strength~$\sigma$.

\section{Discussion}
We have investigated single photon transport properties of chiral and bidirectional waveguides coupled to atomic dimers. We have considered systems of periodically arranged and disordered chains of dimers. Our results may be summarized as follows. The transmission in chiral waveguides is immune to disorder in dimer separations. However, disordered dimer lengths lead to localization, with the localization length depending on the strength of the dipole-dipole interaction. Bidirectional waveguides exhibit an interaction-dependent band structure in the periodic case and localization for both types of disorder.  

We have focused on a specific type of interaction between atoms (namely the dipole-dipole interaction). However, one can also consider a more general situation in which atoms can be coupled through other types of interactions. For instance, in a recent study by Pichler et al. \cite{pichler2015quantum}, spin chains are driven by an on-resonance coherent field, which, under the right conditions, can form dimerized dark states. We leave the investigation of such systems as a direction for future work.

\acknowledgments

This work was supported in part by the NSF grants DMR-1120923 and DMS-1619907.

\bibliography{Paper}

\begin{thebibliography}{44}%
\makeatletter
\providecommand \@ifxundefined [1]{%
 \@ifx{#1\undefined}
}%
\providecommand \@ifnum [1]{%
 \ifnum #1\expandafter \@firstoftwo
 \else \expandafter \@secondoftwo
 \fi
}%
\providecommand \@ifx [1]{%
 \ifx #1\expandafter \@firstoftwo
 \else \expandafter \@secondoftwo
 \fi
}%
\providecommand \natexlab [1]{#1}%
\providecommand \enquote  [1]{``#1''}%
\providecommand \bibnamefont  [1]{#1}%
\providecommand \bibfnamefont [1]{#1}%
\providecommand \citenamefont [1]{#1}%
\providecommand \href@noop [0]{\@secondoftwo}%
\providecommand \href [0]{\begingroup \@sanitize@url \@href}%
\providecommand \@href[1]{\@@startlink{#1}\@@href}%
\providecommand \@@href[1]{\endgroup#1\@@endlink}%
\providecommand \@sanitize@url [0]{\catcode `\\12\catcode `\$12\catcode
  `\&12\catcode `\#12\catcode `\^12\catcode `\_12\catcode `\%12\relax}%
\providecommand \@@startlink[1]{}%
\providecommand \@@endlink[0]{}%
\providecommand \url  [0]{\begingroup\@sanitize@url \@url }%
\providecommand \@url [1]{\endgroup\@href {#1}{\urlprefix }}%
\providecommand \urlprefix  [0]{URL }%
\providecommand \Eprint [0]{\href }%
\providecommand \doibase [0]{http://dx.doi.org/}%
\providecommand \selectlanguage [0]{\@gobble}%
\providecommand \bibinfo  [0]{\@secondoftwo}%
\providecommand \bibfield  [0]{\@secondoftwo}%
\providecommand \translation [1]{[#1]}%
\providecommand \BibitemOpen [0]{}%
\providecommand \bibitemStop [0]{}%
\providecommand \bibitemNoStop [0]{.\EOS\space}%
\providecommand \EOS [0]{\spacefactor3000\relax}%
\providecommand \BibitemShut  [1]{\csname bibitem#1\endcsname}%
\let\auto@bib@innerbib\@empty
%</preamble>
\bibitem [{\citenamefont {Roy}\ \emph {et~al.}(2017)\citenamefont {Roy},
  \citenamefont {Wilson},\ and\ \citenamefont
  {Firstenberg}}]{roy2017colloquium}%
  \BibitemOpen
  \bibfield  {author} {\bibinfo {author} {\bibfnamefont {Dibyendu}\
  \bibnamefont {Roy}}, \bibinfo {author} {\bibfnamefont {CM}~\bibnamefont
  {Wilson}}, \ and\ \bibinfo {author} {\bibfnamefont {Ofer}\ \bibnamefont
  {Firstenberg}},\ }\bibfield  {title} {\enquote {\bibinfo {title} {Colloquium:
  Strongly interacting photons in one-dimensional continuum},}\ }\href@noop {}
  {\bibfield  {journal} {\bibinfo  {journal} {Reviews of Modern Physics}\
  }\textbf {\bibinfo {volume} {89}},\ \bibinfo {pages} {021001} (\bibinfo
  {year} {2017})}\BibitemShut {NoStop}%
\bibitem [{\citenamefont {Hung}\ \emph {et~al.}(2013)\citenamefont {Hung},
  \citenamefont {Meenehan}, \citenamefont {Chang}, \citenamefont {Painter},\
  and\ \citenamefont {Kimble}}]{hung2013trapped}%
  \BibitemOpen
  \bibfield  {author} {\bibinfo {author} {\bibfnamefont {CL}~\bibnamefont
  {Hung}}, \bibinfo {author} {\bibfnamefont {SM}~\bibnamefont {Meenehan}},
  \bibinfo {author} {\bibfnamefont {DE}~\bibnamefont {Chang}}, \bibinfo
  {author} {\bibfnamefont {O}~\bibnamefont {Painter}}, \ and\ \bibinfo {author}
  {\bibfnamefont {HJ}~\bibnamefont {Kimble}},\ }\bibfield  {title} {\enquote
  {\bibinfo {title} {Trapped atoms in one-dimensional photonic crystals},}\
  }\href@noop {} {\bibfield  {journal} {\bibinfo  {journal} {New Journal of
  Physics}\ }\textbf {\bibinfo {volume} {15}},\ \bibinfo {pages} {083026}
  (\bibinfo {year} {2013})}\BibitemShut {NoStop}%
\bibitem [{\citenamefont {Javadi}\ \emph {et~al.}(2015)\citenamefont {Javadi},
  \citenamefont {S{\"o}llner}, \citenamefont {Arcari}, \citenamefont {Hansen},
  \citenamefont {Midolo}, \citenamefont {Mahmoodian}, \citenamefont
  {Kir{\v{s}}ansk{\.e}}, \citenamefont {Pregnolato}, \citenamefont {Lee},
  \citenamefont {Song} \emph {et~al.}}]{javadi2015single}%
  \BibitemOpen
  \bibfield  {author} {\bibinfo {author} {\bibfnamefont {Alisa}\ \bibnamefont
  {Javadi}}, \bibinfo {author} {\bibfnamefont {I}~\bibnamefont {S{\"o}llner}},
  \bibinfo {author} {\bibfnamefont {Marta}\ \bibnamefont {Arcari}}, \bibinfo
  {author} {\bibfnamefont {S~Lindskov}\ \bibnamefont {Hansen}}, \bibinfo
  {author} {\bibfnamefont {Leonardo}\ \bibnamefont {Midolo}}, \bibinfo {author}
  {\bibfnamefont {Sahand}\ \bibnamefont {Mahmoodian}}, \bibinfo {author}
  {\bibfnamefont {G}~\bibnamefont {Kir{\v{s}}ansk{\.e}}}, \bibinfo {author}
  {\bibfnamefont {Tommaso}\ \bibnamefont {Pregnolato}}, \bibinfo {author}
  {\bibfnamefont {EH}~\bibnamefont {Lee}}, \bibinfo {author} {\bibfnamefont
  {JD}~\bibnamefont {Song}},  \emph {et~al.},\ }\bibfield  {title} {\enquote
  {\bibinfo {title} {Single-photon non-linear optics with a quantum dot in a
  waveguide},}\ }\href@noop {} {\bibfield  {journal} {\bibinfo  {journal}
  {Nature communications}\ }\textbf {\bibinfo {volume} {6}},\ \bibinfo {pages}
  {8655} (\bibinfo {year} {2015})}\BibitemShut {NoStop}%
\bibitem [{\citenamefont {Yalla}\ \emph {et~al.}(2014)\citenamefont {Yalla},
  \citenamefont {Sadgrove}, \citenamefont {Nayak},\ and\ \citenamefont
  {Hakuta}}]{yalla2014cavity}%
  \BibitemOpen
  \bibfield  {author} {\bibinfo {author} {\bibfnamefont {Ramachandrarao}\
  \bibnamefont {Yalla}}, \bibinfo {author} {\bibfnamefont {Mark}\ \bibnamefont
  {Sadgrove}}, \bibinfo {author} {\bibfnamefont {Kali~P}\ \bibnamefont
  {Nayak}}, \ and\ \bibinfo {author} {\bibfnamefont {Kohzo}\ \bibnamefont
  {Hakuta}},\ }\bibfield  {title} {\enquote {\bibinfo {title} {Cavity quantum
  electrodynamics on a nanofiber using a composite photonic crystal cavity},}\
  }\href@noop {} {\bibfield  {journal} {\bibinfo  {journal} {Physical review
  letters}\ }\textbf {\bibinfo {volume} {113}},\ \bibinfo {pages} {143601}
  (\bibinfo {year} {2014})}\BibitemShut {NoStop}%
\bibitem [{\citenamefont {Martens}\ \emph {et~al.}(2013)\citenamefont
  {Martens}, \citenamefont {Longo},\ and\ \citenamefont
  {Busch}}]{martens2013photon}%
  \BibitemOpen
  \bibfield  {author} {\bibinfo {author} {\bibfnamefont {Christoph}\
  \bibnamefont {Martens}}, \bibinfo {author} {\bibfnamefont {Paolo}\
  \bibnamefont {Longo}}, \ and\ \bibinfo {author} {\bibfnamefont {Kurt}\
  \bibnamefont {Busch}},\ }\bibfield  {title} {\enquote {\bibinfo {title}
  {Photon transport in one-dimensional systems coupled to three-level quantum
  impurities},}\ }\href@noop {} {\bibfield  {journal} {\bibinfo  {journal} {New
  Journal of Physics}\ }\textbf {\bibinfo {volume} {15}},\ \bibinfo {pages}
  {083019} (\bibinfo {year} {2013})}\BibitemShut {NoStop}%
\bibitem [{\citenamefont {Chang}\ \emph {et~al.}(2014)\citenamefont {Chang},
  \citenamefont {Vuleti{\'c}},\ and\ \citenamefont {Lukin}}]{chang2014quantum}%
  \BibitemOpen
  \bibfield  {author} {\bibinfo {author} {\bibfnamefont {Darrick~E}\
  \bibnamefont {Chang}}, \bibinfo {author} {\bibfnamefont {Vladan}\
  \bibnamefont {Vuleti{\'c}}}, \ and\ \bibinfo {author} {\bibfnamefont
  {Mikhail~D}\ \bibnamefont {Lukin}},\ }\bibfield  {title} {\enquote {\bibinfo
  {title} {Quantum nonlinear optics photon by photon},}\ }\href@noop {}
  {\bibfield  {journal} {\bibinfo  {journal} {Nature Photonics}\ }\textbf
  {\bibinfo {volume} {8}},\ \bibinfo {pages} {685--694} (\bibinfo {year}
  {2014})}\BibitemShut {NoStop}%
\bibitem [{\citenamefont {Fan}\ \emph {et~al.}(2010)\citenamefont {Fan},
  \citenamefont {Kocaba{\c{s}}},\ and\ \citenamefont {Shen}}]{fan2010input}%
  \BibitemOpen
  \bibfield  {author} {\bibinfo {author} {\bibfnamefont {Shanhui}\ \bibnamefont
  {Fan}}, \bibinfo {author} {\bibfnamefont {{\c{S}}{\"u}kr{\"u}~Ekin}\
  \bibnamefont {Kocaba{\c{s}}}}, \ and\ \bibinfo {author} {\bibfnamefont
  {Jung-Tsung}\ \bibnamefont {Shen}},\ }\bibfield  {title} {\enquote {\bibinfo
  {title} {Input-output formalism for few-photon transport in one-dimensional
  nanophotonic waveguides coupled to a qubit},}\ }\href@noop {} {\bibfield
  {journal} {\bibinfo  {journal} {Physical Review A}\ }\textbf {\bibinfo
  {volume} {82}},\ \bibinfo {pages} {063821} (\bibinfo {year}
  {2010})}\BibitemShut {NoStop}%
\bibitem [{\citenamefont {Xu}\ and\ \citenamefont {Fan}(2017)}]{xu2017input}%
  \BibitemOpen
  \bibfield  {author} {\bibinfo {author} {\bibfnamefont {Shanshan}\
  \bibnamefont {Xu}}\ and\ \bibinfo {author} {\bibfnamefont {Shanhui}\
  \bibnamefont {Fan}},\ }\bibfield  {title} {\enquote {\bibinfo {title}
  {Input-output formalism for few-photon transport},}\ }in\ \href@noop {}
  {\emph {\bibinfo {booktitle} {Quantum Plasmonics}}}\ (\bibinfo  {publisher}
  {Springer},\ \bibinfo {year} {2017})\ pp.\ \bibinfo {pages}
  {1--23}\BibitemShut {NoStop}%
\bibitem [{\citenamefont {Shen}\ \emph {et~al.}(2007)\citenamefont {Shen},
  \citenamefont {Povinelli}, \citenamefont {Sandhu},\ and\ \citenamefont
  {Fan}}]{shen2007stopping}%
  \BibitemOpen
  \bibfield  {author} {\bibinfo {author} {\bibfnamefont {Jung-Tsung}\
  \bibnamefont {Shen}}, \bibinfo {author} {\bibfnamefont {ML}~\bibnamefont
  {Povinelli}}, \bibinfo {author} {\bibfnamefont {Sunil}\ \bibnamefont
  {Sandhu}}, \ and\ \bibinfo {author} {\bibfnamefont {Shanhui}\ \bibnamefont
  {Fan}},\ }\bibfield  {title} {\enquote {\bibinfo {title} {Stopping single
  photons in one-dimensional circuit quantum electrodynamics systems},}\
  }\href@noop {} {\bibfield  {journal} {\bibinfo  {journal} {Physical Review
  B}\ }\textbf {\bibinfo {volume} {75}},\ \bibinfo {pages} {035320} (\bibinfo
  {year} {2007})}\BibitemShut {NoStop}%
\bibitem [{\citenamefont {Douglas}\ \emph {et~al.}(2015)\citenamefont
  {Douglas}, \citenamefont {Habibian}, \citenamefont {Hung}, \citenamefont
  {Gorshkov}, \citenamefont {Kimble},\ and\ \citenamefont
  {Chang}}]{douglas2015quantum}%
  \BibitemOpen
  \bibfield  {author} {\bibinfo {author} {\bibfnamefont {James~S}\ \bibnamefont
  {Douglas}}, \bibinfo {author} {\bibfnamefont {H}~\bibnamefont {Habibian}},
  \bibinfo {author} {\bibfnamefont {C-L}\ \bibnamefont {Hung}}, \bibinfo
  {author} {\bibfnamefont {AV}~\bibnamefont {Gorshkov}}, \bibinfo {author}
  {\bibfnamefont {H~Jeff}\ \bibnamefont {Kimble}}, \ and\ \bibinfo {author}
  {\bibfnamefont {Darrick~E}\ \bibnamefont {Chang}},\ }\bibfield  {title}
  {\enquote {\bibinfo {title} {Quantum many-body models with cold atoms coupled
  to photonic crystals},}\ }\href@noop {} {\bibfield  {journal} {\bibinfo
  {journal} {Nature Photonics}\ }\textbf {\bibinfo {volume} {9}},\ \bibinfo
  {pages} {326--331} (\bibinfo {year} {2015})}\BibitemShut {NoStop}%
\bibitem [{\citenamefont {Witthaut}\ and\ \citenamefont
  {S{\o}rensen}(2010)}]{witthaut2010photon}%
  \BibitemOpen
  \bibfield  {author} {\bibinfo {author} {\bibfnamefont {Dirk}\ \bibnamefont
  {Witthaut}}\ and\ \bibinfo {author} {\bibfnamefont {A~S{\o}ndberg}\
  \bibnamefont {S{\o}rensen}},\ }\bibfield  {title} {\enquote {\bibinfo {title}
  {Photon scattering by a three-level emitter in a one-dimensional
  waveguide},}\ }\href@noop {} {\bibfield  {journal} {\bibinfo  {journal} {New
  Journal of Physics}\ }\textbf {\bibinfo {volume} {12}},\ \bibinfo {pages}
  {043052} (\bibinfo {year} {2010})}\BibitemShut {NoStop}%
\bibitem [{\citenamefont {Mirza}\ \emph {et~al.}(2017)\citenamefont {Mirza},
  \citenamefont {Hoskins},\ and\ \citenamefont
  {Schotland}}]{mirza2017chirality}%
  \BibitemOpen
  \bibfield  {author} {\bibinfo {author} {\bibfnamefont {Imran~M}\ \bibnamefont
  {Mirza}}, \bibinfo {author} {\bibfnamefont {Jeremy~G}\ \bibnamefont
  {Hoskins}}, \ and\ \bibinfo {author} {\bibfnamefont {John~C}\ \bibnamefont
  {Schotland}},\ }\bibfield  {title} {\enquote {\bibinfo {title} {Chirality,
  band structure, and localization in waveguide quantum electrodynamics},}\
  }\href@noop {} {\bibfield  {journal} {\bibinfo  {journal} {Physical Review
  A}\ }\textbf {\bibinfo {volume} {96}},\ \bibinfo {pages} {053804} (\bibinfo
  {year} {2017})}\BibitemShut {NoStop}%
\bibitem [{\citenamefont {Mirza}\ and\ \citenamefont
  {Schotland}(2018)}]{mirza2017influence}%
  \BibitemOpen
  \bibfield  {author} {\bibinfo {author} {\bibfnamefont {Imran~M}\ \bibnamefont
  {Mirza}}\ and\ \bibinfo {author} {\bibfnamefont {John~C}\ \bibnamefont
  {Schotland}},\ }\bibfield  {title} {\enquote {\bibinfo {title} {Influence of
  disorder on electromagnetically induced transparency in chiral waveguide
  quantum electrodynamics},}\ }\href@noop {} {\bibfield  {journal} {\bibinfo
  {journal} {JOSA B}\ }\textbf {\bibinfo {volume} {35}},\ \bibinfo {pages}
  {1149--1158} (\bibinfo {year} {2018})}\BibitemShut {NoStop}%
\bibitem [{\citenamefont {Kocaba{\c{s}}}\ \emph {et~al.}(2012)\citenamefont
  {Kocaba{\c{s}}}, \citenamefont {Rephaeli},\ and\ \citenamefont
  {Fan}}]{kocabacs2012resonance}%
  \BibitemOpen
  \bibfield  {author} {\bibinfo {author} {\bibfnamefont
  {{\c{S}}{\"u}kr{\"u}~Ekin}\ \bibnamefont {Kocaba{\c{s}}}}, \bibinfo {author}
  {\bibfnamefont {Eden}\ \bibnamefont {Rephaeli}}, \ and\ \bibinfo {author}
  {\bibfnamefont {Shanhui}\ \bibnamefont {Fan}},\ }\bibfield  {title} {\enquote
  {\bibinfo {title} {Resonance fluorescence in a waveguide geometry},}\
  }\href@noop {} {\bibfield  {journal} {\bibinfo  {journal} {Physical Review
  A}\ }\textbf {\bibinfo {volume} {85}},\ \bibinfo {pages} {023817} (\bibinfo
  {year} {2012})}\BibitemShut {NoStop}%
\bibitem [{\citenamefont {Gonzalez-Ballestero}\ \emph
  {et~al.}(2015)\citenamefont {Gonzalez-Ballestero}, \citenamefont
  {Gonzalez-Tudela}, \citenamefont {Garcia-Vidal},\ and\ \citenamefont
  {Moreno}}]{gonzalez2015chiral}%
  \BibitemOpen
  \bibfield  {author} {\bibinfo {author} {\bibfnamefont {Carlos}\ \bibnamefont
  {Gonzalez-Ballestero}}, \bibinfo {author} {\bibfnamefont {Alejandro}\
  \bibnamefont {Gonzalez-Tudela}}, \bibinfo {author} {\bibfnamefont
  {Francisco~J}\ \bibnamefont {Garcia-Vidal}}, \ and\ \bibinfo {author}
  {\bibfnamefont {Esteban}\ \bibnamefont {Moreno}},\ }\bibfield  {title}
  {\enquote {\bibinfo {title} {Chiral route to spontaneous entanglement
  generation},}\ }\href@noop {} {\bibfield  {journal} {\bibinfo  {journal}
  {Physical Review B}\ }\textbf {\bibinfo {volume} {92}},\ \bibinfo {pages}
  {155304} (\bibinfo {year} {2015})}\BibitemShut {NoStop}%
\bibitem [{\citenamefont {Mirza}\ and\ \citenamefont
  {Schotland}(2016{\natexlab{a}})}]{mirza2016multiqubit}%
  \BibitemOpen
  \bibfield  {author} {\bibinfo {author} {\bibfnamefont {Imran~M}\ \bibnamefont
  {Mirza}}\ and\ \bibinfo {author} {\bibfnamefont {John~C}\ \bibnamefont
  {Schotland}},\ }\bibfield  {title} {\enquote {\bibinfo {title} {Multiqubit
  entanglement in bidirectional-chiral-waveguide qed},}\ }\href@noop {}
  {\bibfield  {journal} {\bibinfo  {journal} {Physical Review A}\ }\textbf
  {\bibinfo {volume} {94}},\ \bibinfo {pages} {012302} (\bibinfo {year}
  {2016}{\natexlab{a}})}\BibitemShut {NoStop}%
\bibitem [{\citenamefont {Mirza}\ and\ \citenamefont
  {Schotland}(2016{\natexlab{b}})}]{mirza2016two}%
  \BibitemOpen
  \bibfield  {author} {\bibinfo {author} {\bibfnamefont {Imran~M}\ \bibnamefont
  {Mirza}}\ and\ \bibinfo {author} {\bibfnamefont {John~C}\ \bibnamefont
  {Schotland}},\ }\bibfield  {title} {\enquote {\bibinfo {title} {Two-photon
  entanglement in multiqubit bidirectional-waveguide qed},}\ }\href@noop {}
  {\bibfield  {journal} {\bibinfo  {journal} {Physical Review A}\ }\textbf
  {\bibinfo {volume} {94}},\ \bibinfo {pages} {012309} (\bibinfo {year}
  {2016}{\natexlab{b}})}\BibitemShut {NoStop}%
\bibitem [{\citenamefont {Coles}\ \emph {et~al.}(2016)\citenamefont {Coles},
  \citenamefont {Price}, \citenamefont {Dixon}, \citenamefont {Royall},
  \citenamefont {Clarke}, \citenamefont {Kok}, \citenamefont {Skolnick},
  \citenamefont {Fox},\ and\ \citenamefont {Makhonin}}]{coles2016chirality}%
  \BibitemOpen
  \bibfield  {author} {\bibinfo {author} {\bibfnamefont {RJ}~\bibnamefont
  {Coles}}, \bibinfo {author} {\bibfnamefont {DM}~\bibnamefont {Price}},
  \bibinfo {author} {\bibfnamefont {JE}~\bibnamefont {Dixon}}, \bibinfo
  {author} {\bibfnamefont {B}~\bibnamefont {Royall}}, \bibinfo {author}
  {\bibfnamefont {E}~\bibnamefont {Clarke}}, \bibinfo {author} {\bibfnamefont
  {P}~\bibnamefont {Kok}}, \bibinfo {author} {\bibfnamefont {MS}~\bibnamefont
  {Skolnick}}, \bibinfo {author} {\bibfnamefont {AM}~\bibnamefont {Fox}}, \
  and\ \bibinfo {author} {\bibfnamefont {MN}~\bibnamefont {Makhonin}},\
  }\bibfield  {title} {\enquote {\bibinfo {title} {Chirality of nanophotonic
  waveguide with embedded quantum emitter for unidirectional spin transfer},}\
  }\href@noop {} {\bibfield  {journal} {\bibinfo  {journal} {Nature
  communications}\ }\textbf {\bibinfo {volume} {7}} (\bibinfo {year}
  {2016})}\BibitemShut {NoStop}%
\bibitem [{\citenamefont {Mitsch}\ \emph {et~al.}(2014)\citenamefont {Mitsch},
  \citenamefont {Sayrin}, \citenamefont {Albrecht}, \citenamefont
  {Schneeweiss},\ and\ \citenamefont {Rauschenbeutel}}]{mitsch2014quantum}%
  \BibitemOpen
  \bibfield  {author} {\bibinfo {author} {\bibfnamefont {R}~\bibnamefont
  {Mitsch}}, \bibinfo {author} {\bibfnamefont {C}~\bibnamefont {Sayrin}},
  \bibinfo {author} {\bibfnamefont {B}~\bibnamefont {Albrecht}}, \bibinfo
  {author} {\bibfnamefont {P}~\bibnamefont {Schneeweiss}}, \ and\ \bibinfo
  {author} {\bibfnamefont {A}~\bibnamefont {Rauschenbeutel}},\ }\bibfield
  {title} {\enquote {\bibinfo {title} {Quantum state-controlled directional
  spontaneous emission of photons into a nanophotonic waveguide},}\ }\href@noop
  {} {\bibfield  {journal} {\bibinfo  {journal} {Nature communications}\
  }\textbf {\bibinfo {volume} {5}} (\bibinfo {year} {2014})}\BibitemShut
  {NoStop}%
\bibitem [{\citenamefont {Tiecke}\ \emph {et~al.}(2014)\citenamefont {Tiecke},
  \citenamefont {Thompson}, \citenamefont {de~Leon}, \citenamefont {Liu},
  \citenamefont {Vuletic},\ and\ \citenamefont
  {Lukin}}]{tiecke2014nanophotonic}%
  \BibitemOpen
  \bibfield  {author} {\bibinfo {author} {\bibfnamefont {TG}~\bibnamefont
  {Tiecke}}, \bibinfo {author} {\bibfnamefont {JD}~\bibnamefont {Thompson}},
  \bibinfo {author} {\bibfnamefont {NP}~\bibnamefont {de~Leon}}, \bibinfo
  {author} {\bibfnamefont {LR}~\bibnamefont {Liu}}, \bibinfo {author}
  {\bibfnamefont {V}~\bibnamefont {Vuletic}}, \ and\ \bibinfo {author}
  {\bibfnamefont {MD}~\bibnamefont {Lukin}},\ }\bibfield  {title} {\enquote
  {\bibinfo {title} {Nanophotonic quantum phase switch with a single atom},}\
  }\href@noop {} {\bibfield  {journal} {\bibinfo  {journal} {Nature}\ }\textbf
  {\bibinfo {volume} {508}},\ \bibinfo {pages} {241--245} (\bibinfo {year}
  {2014})}\BibitemShut {NoStop}%
\bibitem [{\citenamefont {Paulisch}\ \emph {et~al.}(2016)\citenamefont
  {Paulisch}, \citenamefont {Kimble},\ and\ \citenamefont
  {Gonz{\'a}lez-Tudela}}]{paulisch2016universal}%
  \BibitemOpen
  \bibfield  {author} {\bibinfo {author} {\bibfnamefont {Vanessa}\ \bibnamefont
  {Paulisch}}, \bibinfo {author} {\bibfnamefont {HJ}~\bibnamefont {Kimble}}, \
  and\ \bibinfo {author} {\bibfnamefont {Alejandro}\ \bibnamefont
  {Gonz{\'a}lez-Tudela}},\ }\bibfield  {title} {\enquote {\bibinfo {title}
  {Universal quantum computation in waveguide qed using decoherence free
  subspaces},}\ }\href@noop {} {\bibfield  {journal} {\bibinfo  {journal} {New
  Journal of Physics}\ }\textbf {\bibinfo {volume} {18}},\ \bibinfo {pages}
  {043041} (\bibinfo {year} {2016})}\BibitemShut {NoStop}%
\bibitem [{\citenamefont {Lodahl}\ \emph {et~al.}(2017)\citenamefont {Lodahl},
  \citenamefont {Mahmoodian}, \citenamefont {Stobbe}, \citenamefont
  {Rauschenbeutel}, \citenamefont {Schneeweiss}, \citenamefont {Volz},
  \citenamefont {Pichler},\ and\ \citenamefont {Zoller}}]{lodahl2017chiral}%
  \BibitemOpen
  \bibfield  {author} {\bibinfo {author} {\bibfnamefont {Peter}\ \bibnamefont
  {Lodahl}}, \bibinfo {author} {\bibfnamefont {Sahand}\ \bibnamefont
  {Mahmoodian}}, \bibinfo {author} {\bibfnamefont {S{\o}ren}\ \bibnamefont
  {Stobbe}}, \bibinfo {author} {\bibfnamefont {Arno}\ \bibnamefont
  {Rauschenbeutel}}, \bibinfo {author} {\bibfnamefont {Philipp}\ \bibnamefont
  {Schneeweiss}}, \bibinfo {author} {\bibfnamefont {J{\"u}rgen}\ \bibnamefont
  {Volz}}, \bibinfo {author} {\bibfnamefont {Hannes}\ \bibnamefont {Pichler}},
  \ and\ \bibinfo {author} {\bibfnamefont {Peter}\ \bibnamefont {Zoller}},\
  }\bibfield  {title} {\enquote {\bibinfo {title} {Chiral quantum optics},}\
  }\href@noop {} {\bibfield  {journal} {\bibinfo  {journal} {Nature}\ }\textbf
  {\bibinfo {volume} {541}},\ \bibinfo {pages} {473--480} (\bibinfo {year}
  {2017})}\BibitemShut {NoStop}%
\bibitem [{\citenamefont {Milonni}\ and\ \citenamefont
  {Knight}(1974)}]{milonni1974retardation}%
  \BibitemOpen
  \bibfield  {author} {\bibinfo {author} {\bibfnamefont {PW}~\bibnamefont
  {Milonni}}\ and\ \bibinfo {author} {\bibfnamefont {Pl~L}\ \bibnamefont
  {Knight}},\ }\bibfield  {title} {\enquote {\bibinfo {title} {Retardation in
  the resonant interaction of two identical atoms},}\ }\href@noop {} {\bibfield
   {journal} {\bibinfo  {journal} {Physical Review A}\ }\textbf {\bibinfo
  {volume} {10}},\ \bibinfo {pages} {1096} (\bibinfo {year}
  {1974})}\BibitemShut {NoStop}%
\bibitem [{\citenamefont {Scully}\ and\ \citenamefont
  {Svidzinsky}(2009)}]{scully2009super}%
  \BibitemOpen
  \bibfield  {author} {\bibinfo {author} {\bibfnamefont {Marlan~O}\
  \bibnamefont {Scully}}\ and\ \bibinfo {author} {\bibfnamefont {Anatoly~A}\
  \bibnamefont {Svidzinsky}},\ }\bibfield  {title} {\enquote {\bibinfo {title}
  {The super of superradiance},}\ }\href@noop {} {\bibfield  {journal}
  {\bibinfo  {journal} {Science}\ }\textbf {\bibinfo {volume} {325}},\ \bibinfo
  {pages} {1510--1511} (\bibinfo {year} {2009})}\BibitemShut {NoStop}%
\bibitem [{\citenamefont {Roof}\ \emph {et~al.}(2016)\citenamefont {Roof},
  \citenamefont {Kemp}, \citenamefont {Havey},\ and\ \citenamefont
  {Sokolov}}]{roof2016observation}%
  \BibitemOpen
  \bibfield  {author} {\bibinfo {author} {\bibfnamefont {SJ}~\bibnamefont
  {Roof}}, \bibinfo {author} {\bibfnamefont {KJ}~\bibnamefont {Kemp}}, \bibinfo
  {author} {\bibfnamefont {MD}~\bibnamefont {Havey}}, \ and\ \bibinfo {author}
  {\bibfnamefont {IM}~\bibnamefont {Sokolov}},\ }\bibfield  {title} {\enquote
  {\bibinfo {title} {Observation of single-photon superradiance and the
  cooperative lamb shift in an extended sample of cold atoms},}\ }\href@noop {}
  {\bibfield  {journal} {\bibinfo  {journal} {Physical review letters}\
  }\textbf {\bibinfo {volume} {117}},\ \bibinfo {pages} {073003} (\bibinfo
  {year} {2016})}\BibitemShut {NoStop}%
\bibitem [{\citenamefont {Zheng}\ and\ \citenamefont
  {Baranger}(2013)}]{zheng2013persistent}%
  \BibitemOpen
  \bibfield  {author} {\bibinfo {author} {\bibfnamefont {Huaixiu}\ \bibnamefont
  {Zheng}}\ and\ \bibinfo {author} {\bibfnamefont {Harold~U}\ \bibnamefont
  {Baranger}},\ }\bibfield  {title} {\enquote {\bibinfo {title} {Persistent
  quantum beats and long-distance entanglement from waveguide-mediated
  interactions},}\ }\href@noop {} {\bibfield  {journal} {\bibinfo  {journal}
  {Physical review letters}\ }\textbf {\bibinfo {volume} {110}},\ \bibinfo
  {pages} {113601} (\bibinfo {year} {2013})}\BibitemShut {NoStop}%
\bibitem [{\citenamefont {Facchi}\ \emph {et~al.}(2016)\citenamefont {Facchi},
  \citenamefont {Kim}, \citenamefont {Pascazio}, \citenamefont {Pepe},
  \citenamefont {Pomarico},\ and\ \citenamefont {Tufarelli}}]{facchi2016bound}%
  \BibitemOpen
  \bibfield  {author} {\bibinfo {author} {\bibfnamefont {Paolo}\ \bibnamefont
  {Facchi}}, \bibinfo {author} {\bibfnamefont {MS}~\bibnamefont {Kim}},
  \bibinfo {author} {\bibfnamefont {Saverio}\ \bibnamefont {Pascazio}},
  \bibinfo {author} {\bibfnamefont {Francesco~V}\ \bibnamefont {Pepe}},
  \bibinfo {author} {\bibfnamefont {Domenico}\ \bibnamefont {Pomarico}}, \ and\
  \bibinfo {author} {\bibfnamefont {Tommaso}\ \bibnamefont {Tufarelli}},\
  }\bibfield  {title} {\enquote {\bibinfo {title} {Bound states and
  entanglement generation in waveguide quantum electrodynamics},}\ }\href@noop
  {} {\bibfield  {journal} {\bibinfo  {journal} {Physical Review A}\ }\textbf
  {\bibinfo {volume} {94}},\ \bibinfo {pages} {043839} (\bibinfo {year}
  {2016})}\BibitemShut {NoStop}%
\bibitem [{\citenamefont {Corzo}\ \emph {et~al.}(2016)\citenamefont {Corzo},
  \citenamefont {Gouraud}, \citenamefont {Chandra}, \citenamefont {Goban},
  \citenamefont {Sheremet}, \citenamefont {Kupriyanov},\ and\ \citenamefont
  {Laurat}}]{corzo2016large}%
  \BibitemOpen
  \bibfield  {author} {\bibinfo {author} {\bibfnamefont {Neil~V}\ \bibnamefont
  {Corzo}}, \bibinfo {author} {\bibfnamefont {Baptiste}\ \bibnamefont
  {Gouraud}}, \bibinfo {author} {\bibfnamefont {Aveek}\ \bibnamefont
  {Chandra}}, \bibinfo {author} {\bibfnamefont {Akihisa}\ \bibnamefont
  {Goban}}, \bibinfo {author} {\bibfnamefont {Alexandra~S}\ \bibnamefont
  {Sheremet}}, \bibinfo {author} {\bibfnamefont {Dmitriy~V}\ \bibnamefont
  {Kupriyanov}}, \ and\ \bibinfo {author} {\bibfnamefont {Julien}\ \bibnamefont
  {Laurat}},\ }\bibfield  {title} {\enquote {\bibinfo {title} {Large bragg
  reflection from one-dimensional chains of trapped atoms near a nanoscale
  waveguide},}\ }\href@noop {} {\bibfield  {journal} {\bibinfo  {journal}
  {Physical review letters}\ }\textbf {\bibinfo {volume} {117}},\ \bibinfo
  {pages} {133603} (\bibinfo {year} {2016})}\BibitemShut {NoStop}%
\bibitem [{\citenamefont {Agarwal}(2013)}]{agarwal2013quantum}%
  \BibitemOpen
  \bibfield  {author} {\bibinfo {author} {\bibfnamefont {Girish~S}\
  \bibnamefont {Agarwal}},\ }\href@noop {} {\emph {\bibinfo {title} {Quantum
  Optics}}}\ (\bibinfo  {publisher} {Cambridge University Press},\ \bibinfo
  {year} {2013})\BibitemShut {NoStop}%
\bibitem [{\citenamefont {Cheng}\ \emph {et~al.}(2017)\citenamefont {Cheng},
  \citenamefont {Xu},\ and\ \citenamefont {Agarwal}}]{cheng2017waveguide}%
  \BibitemOpen
  \bibfield  {author} {\bibinfo {author} {\bibfnamefont {Mu-Tian}\ \bibnamefont
  {Cheng}}, \bibinfo {author} {\bibfnamefont {Jingping}\ \bibnamefont {Xu}}, \
  and\ \bibinfo {author} {\bibfnamefont {Girish~S}\ \bibnamefont {Agarwal}},\
  }\bibfield  {title} {\enquote {\bibinfo {title} {Waveguide transport mediated
  by strong coupling with atoms},}\ }\href@noop {} {\bibfield  {journal}
  {\bibinfo  {journal} {Physical Review A}\ }\textbf {\bibinfo {volume} {95}},\
  \bibinfo {pages} {053807} (\bibinfo {year} {2017})}\BibitemShut {NoStop}%
\bibitem [{\citenamefont {Marcuzzi}\ \emph {et~al.}(2017)\citenamefont
  {Marcuzzi}, \citenamefont {Min{\'a}{\v{r}}}, \citenamefont {Barredo},
  \citenamefont {de~L{\'e}s{\'e}leuc}, \citenamefont {Labuhn}, \citenamefont
  {Lahaye}, \citenamefont {Browaeys}, \citenamefont {Levi},\ and\ \citenamefont
  {Lesanovsky}}]{marcuzzi2017facilitation}%
  \BibitemOpen
  \bibfield  {author} {\bibinfo {author} {\bibfnamefont {Matteo}\ \bibnamefont
  {Marcuzzi}}, \bibinfo {author} {\bibfnamefont {Ji{\v{r}}{\'\i}}\ \bibnamefont
  {Min{\'a}{\v{r}}}}, \bibinfo {author} {\bibfnamefont {Daniel}\ \bibnamefont
  {Barredo}}, \bibinfo {author} {\bibfnamefont {Sylvain}\ \bibnamefont
  {de~L{\'e}s{\'e}leuc}}, \bibinfo {author} {\bibfnamefont {Henning}\
  \bibnamefont {Labuhn}}, \bibinfo {author} {\bibfnamefont {Thierry}\
  \bibnamefont {Lahaye}}, \bibinfo {author} {\bibfnamefont {Antoine}\
  \bibnamefont {Browaeys}}, \bibinfo {author} {\bibfnamefont {Emanuele}\
  \bibnamefont {Levi}}, \ and\ \bibinfo {author} {\bibfnamefont {Igor}\
  \bibnamefont {Lesanovsky}},\ }\bibfield  {title} {\enquote {\bibinfo {title}
  {Facilitation dynamics and localization phenomena in rydberg lattice gases
  with position disorder},}\ }\href@noop {} {\bibfield  {journal} {\bibinfo
  {journal} {Physical Review Letters}\ }\textbf {\bibinfo {volume} {118}},\
  \bibinfo {pages} {063606} (\bibinfo {year} {2017})}\BibitemShut {NoStop}%
\bibitem [{\citenamefont {Vermersch}\ \emph {et~al.}(2016)\citenamefont
  {Vermersch}, \citenamefont {Ramos}, \citenamefont {Hauke},\ and\
  \citenamefont {Zoller}}]{vermersch2016implementation}%
  \BibitemOpen
  \bibfield  {author} {\bibinfo {author} {\bibfnamefont {Beno{\^\i}t}\
  \bibnamefont {Vermersch}}, \bibinfo {author} {\bibfnamefont {Tom{\'a}s}\
  \bibnamefont {Ramos}}, \bibinfo {author} {\bibfnamefont {Philipp}\
  \bibnamefont {Hauke}}, \ and\ \bibinfo {author} {\bibfnamefont {Peter}\
  \bibnamefont {Zoller}},\ }\bibfield  {title} {\enquote {\bibinfo {title}
  {Implementation of chiral quantum optics with rydberg and trapped-ion
  setups},}\ }\href@noop {} {\bibfield  {journal} {\bibinfo  {journal}
  {Physical Review A}\ }\textbf {\bibinfo {volume} {93}},\ \bibinfo {pages}
  {063830} (\bibinfo {year} {2016})}\BibitemShut {NoStop}%
\bibitem [{\citenamefont {Glaetzle}\ \emph {et~al.}(2014)\citenamefont
  {Glaetzle}, \citenamefont {Dalmonte}, \citenamefont {Nath}, \citenamefont
  {Rousochatzakis}, \citenamefont {Moessner},\ and\ \citenamefont
  {Zoller}}]{glaetzle2014quantum}%
  \BibitemOpen
  \bibfield  {author} {\bibinfo {author} {\bibfnamefont {Alexander~W}\
  \bibnamefont {Glaetzle}}, \bibinfo {author} {\bibfnamefont {Marcello}\
  \bibnamefont {Dalmonte}}, \bibinfo {author} {\bibfnamefont {Rejish}\
  \bibnamefont {Nath}}, \bibinfo {author} {\bibfnamefont {Ioannis}\
  \bibnamefont {Rousochatzakis}}, \bibinfo {author} {\bibfnamefont {Roderich}\
  \bibnamefont {Moessner}}, \ and\ \bibinfo {author} {\bibfnamefont {Peter}\
  \bibnamefont {Zoller}},\ }\bibfield  {title} {\enquote {\bibinfo {title}
  {Quantum spin-ice and dimer models with rydberg atoms},}\ }\href@noop {}
  {\bibfield  {journal} {\bibinfo  {journal} {Physical Review X}\ }\textbf
  {\bibinfo {volume} {4}},\ \bibinfo {pages} {041037} (\bibinfo {year}
  {2014})}\BibitemShut {NoStop}%
\bibitem [{\citenamefont {Hofmann}\ \emph {et~al.}(2014)\citenamefont
  {Hofmann}, \citenamefont {G{\"u}nter}, \citenamefont {Schempp}, \citenamefont
  {M{\"u}ller}, \citenamefont {Faber}, \citenamefont {Busche}, \citenamefont
  {Robert-de Saint-Vincent}, \citenamefont {Whitlock},\ and\ \citenamefont
  {Weidem{\"u}ller}}]{hofmann2014experimental}%
  \BibitemOpen
  \bibfield  {author} {\bibinfo {author} {\bibfnamefont {CS}~\bibnamefont
  {Hofmann}}, \bibinfo {author} {\bibfnamefont {G}~\bibnamefont {G{\"u}nter}},
  \bibinfo {author} {\bibfnamefont {H}~\bibnamefont {Schempp}}, \bibinfo
  {author} {\bibfnamefont {Nele~LM}\ \bibnamefont {M{\"u}ller}}, \bibinfo
  {author} {\bibfnamefont {A}~\bibnamefont {Faber}}, \bibinfo {author}
  {\bibfnamefont {H}~\bibnamefont {Busche}}, \bibinfo {author} {\bibfnamefont
  {M}~\bibnamefont {Robert-de Saint-Vincent}}, \bibinfo {author} {\bibfnamefont
  {S}~\bibnamefont {Whitlock}}, \ and\ \bibinfo {author} {\bibfnamefont
  {M}~\bibnamefont {Weidem{\"u}ller}},\ }\bibfield  {title} {\enquote {\bibinfo
  {title} {An experimental approach for investigating many-body phenomena in
  rydberg-interacting quantum systems},}\ }\href@noop {} {\bibfield  {journal}
  {\bibinfo  {journal} {Frontiers of Physics}\ }\textbf {\bibinfo {volume}
  {9}},\ \bibinfo {pages} {571--586} (\bibinfo {year} {2014})}\BibitemShut
  {NoStop}%
\bibitem [{\citenamefont {Fu}\ \emph {et~al.}(2008)\citenamefont {Fu},
  \citenamefont {Santori}, \citenamefont {Barclay}, \citenamefont
  {Aharonovich}, \citenamefont {Prawer}, \citenamefont {Meyer}, \citenamefont
  {Holm},\ and\ \citenamefont {Beausoleil}}]{fu2008coupling}%
  \BibitemOpen
  \bibfield  {author} {\bibinfo {author} {\bibfnamefont {K-MC}\ \bibnamefont
  {Fu}}, \bibinfo {author} {\bibfnamefont {C}~\bibnamefont {Santori}}, \bibinfo
  {author} {\bibfnamefont {PE}~\bibnamefont {Barclay}}, \bibinfo {author}
  {\bibfnamefont {I}~\bibnamefont {Aharonovich}}, \bibinfo {author}
  {\bibfnamefont {S}~\bibnamefont {Prawer}}, \bibinfo {author} {\bibfnamefont
  {N}~\bibnamefont {Meyer}}, \bibinfo {author} {\bibfnamefont {AM}~\bibnamefont
  {Holm}}, \ and\ \bibinfo {author} {\bibfnamefont {RG}~\bibnamefont
  {Beausoleil}},\ }\bibfield  {title} {\enquote {\bibinfo {title} {Coupling of
  nitrogen-vacancy centers in diamond to a gap waveguide},}\ }\href@noop {}
  {\bibfield  {journal} {\bibinfo  {journal} {Applied Physics Letters}\
  }\textbf {\bibinfo {volume} {93}},\ \bibinfo {pages} {234107} (\bibinfo
  {year} {2008})}\BibitemShut {NoStop}%
\bibitem [{\citenamefont {Hattermann}\ \emph {et~al.}(2017)\citenamefont
  {Hattermann}, \citenamefont {Bothner}, \citenamefont {Ley}, \citenamefont
  {Ferdinand}, \citenamefont {Wiedmaier}, \citenamefont {S{\'a}rk{\'a}ny},
  \citenamefont {Kleiner}, \citenamefont {Koelle},\ and\ \citenamefont
  {Fort{\'a}gh}}]{hattermann2017coupling}%
  \BibitemOpen
  \bibfield  {author} {\bibinfo {author} {\bibfnamefont {H}~\bibnamefont
  {Hattermann}}, \bibinfo {author} {\bibfnamefont {D}~\bibnamefont {Bothner}},
  \bibinfo {author} {\bibfnamefont {LY}~\bibnamefont {Ley}}, \bibinfo {author}
  {\bibfnamefont {B}~\bibnamefont {Ferdinand}}, \bibinfo {author}
  {\bibfnamefont {D}~\bibnamefont {Wiedmaier}}, \bibinfo {author}
  {\bibfnamefont {L}~\bibnamefont {S{\'a}rk{\'a}ny}}, \bibinfo {author}
  {\bibfnamefont {R}~\bibnamefont {Kleiner}}, \bibinfo {author} {\bibfnamefont
  {D}~\bibnamefont {Koelle}}, \ and\ \bibinfo {author} {\bibfnamefont
  {J}~\bibnamefont {Fort{\'a}gh}},\ }\bibfield  {title} {\enquote {\bibinfo
  {title} {Coupling ultracold atoms to a superconducting coplanar waveguide
  resonator},}\ }\href@noop {} {\bibfield  {journal} {\bibinfo  {journal}
  {Nature communications}\ }\textbf {\bibinfo {volume} {8}},\ \bibinfo {pages}
  {2254} (\bibinfo {year} {2017})}\BibitemShut {NoStop}%
\bibitem [{\citenamefont {Lalumiere}\ \emph {et~al.}(2013)\citenamefont
  {Lalumiere}, \citenamefont {Sanders}, \citenamefont {van Loo}, \citenamefont
  {Fedorov}, \citenamefont {Wallraff},\ and\ \citenamefont
  {Blais}}]{lalumiere2013input}%
  \BibitemOpen
  \bibfield  {author} {\bibinfo {author} {\bibfnamefont {Kevin}\ \bibnamefont
  {Lalumiere}}, \bibinfo {author} {\bibfnamefont {Barry~C}\ \bibnamefont
  {Sanders}}, \bibinfo {author} {\bibfnamefont {Arjan~F}\ \bibnamefont {van
  Loo}}, \bibinfo {author} {\bibfnamefont {Arkady}\ \bibnamefont {Fedorov}},
  \bibinfo {author} {\bibfnamefont {Andreas}\ \bibnamefont {Wallraff}}, \ and\
  \bibinfo {author} {\bibfnamefont {Alexandre}\ \bibnamefont {Blais}},\
  }\bibfield  {title} {\enquote {\bibinfo {title} {Input-output theory for
  waveguide qed with an ensemble of inhomogeneous atoms},}\ }\href@noop {}
  {\bibfield  {journal} {\bibinfo  {journal} {arXiv preprint arXiv:1305.7135}\
  } (\bibinfo {year} {2013})}\BibitemShut {NoStop}%
\bibitem [{\citenamefont {Bozhevolnyi}\ \emph {et~al.}(2017)\citenamefont
  {Bozhevolnyi}, \citenamefont {Martin-Moreno},\ and\ \citenamefont
  {Garcia-Vidal}}]{bozhevolnyi2017quantum}%
  \BibitemOpen
  \bibfield  {author} {\bibinfo {author} {\bibfnamefont {Sergey~I}\
  \bibnamefont {Bozhevolnyi}}, \bibinfo {author} {\bibfnamefont {Luis}\
  \bibnamefont {Martin-Moreno}}, \ and\ \bibinfo {author} {\bibfnamefont
  {Francisco}\ \bibnamefont {Garcia-Vidal}},\ }\href@noop {} {\emph {\bibinfo
  {title} {Quantum Plasmonics}}}\ (\bibinfo  {publisher} {Springer},\ \bibinfo
  {year} {2017})\BibitemShut {NoStop}%
\bibitem [{\citenamefont {Akimov}\ \emph {et~al.}(2007)\citenamefont {Akimov},
  \citenamefont {Mukherjee}, \citenamefont {Yu}, \citenamefont {Chang},
  \citenamefont {Zibrov}, \citenamefont {Hemmer}, \citenamefont {Park},\ and\
  \citenamefont {Lukin}}]{akimov2007generation}%
  \BibitemOpen
  \bibfield  {author} {\bibinfo {author} {\bibfnamefont {AV}~\bibnamefont
  {Akimov}}, \bibinfo {author} {\bibfnamefont {A}~\bibnamefont {Mukherjee}},
  \bibinfo {author} {\bibfnamefont {CL}~\bibnamefont {Yu}}, \bibinfo {author}
  {\bibfnamefont {DE}~\bibnamefont {Chang}}, \bibinfo {author} {\bibfnamefont
  {AS}~\bibnamefont {Zibrov}}, \bibinfo {author} {\bibfnamefont
  {PR}~\bibnamefont {Hemmer}}, \bibinfo {author} {\bibfnamefont
  {H}~\bibnamefont {Park}}, \ and\ \bibinfo {author} {\bibfnamefont
  {MD}~\bibnamefont {Lukin}},\ }\bibfield  {title} {\enquote {\bibinfo {title}
  {Generation of single optical plasmons in metallic nanowires coupled to
  quantum dots},}\ }\href@noop {} {\bibfield  {journal} {\bibinfo  {journal}
  {Nature}\ }\textbf {\bibinfo {volume} {450}},\ \bibinfo {pages} {402--406}
  (\bibinfo {year} {2007})}\BibitemShut {NoStop}%
\bibitem [{\citenamefont {Chang}\ \emph {et~al.}(2006)\citenamefont {Chang},
  \citenamefont {S{\o}rensen}, \citenamefont {Hemmer},\ and\ \citenamefont
  {Lukin}}]{chang2006quantum}%
  \BibitemOpen
  \bibfield  {author} {\bibinfo {author} {\bibfnamefont {DE}~\bibnamefont
  {Chang}}, \bibinfo {author} {\bibfnamefont {Anders~S{\o}ndberg}\ \bibnamefont
  {S{\o}rensen}}, \bibinfo {author} {\bibfnamefont {PR}~\bibnamefont {Hemmer}},
  \ and\ \bibinfo {author} {\bibfnamefont {MD}~\bibnamefont {Lukin}},\
  }\bibfield  {title} {\enquote {\bibinfo {title} {Quantum optics with surface
  plasmons},}\ }\href@noop {} {\bibfield  {journal} {\bibinfo  {journal}
  {Physical review letters}\ }\textbf {\bibinfo {volume} {97}},\ \bibinfo
  {pages} {053002} (\bibinfo {year} {2006})}\BibitemShut {NoStop}%
\bibitem [{\citenamefont {Markos}\ and\ \citenamefont
  {Soukoulis}(2008)}]{markos2008wave}%
  \BibitemOpen
  \bibfield  {author} {\bibinfo {author} {\bibfnamefont {Peter}\ \bibnamefont
  {Markos}}\ and\ \bibinfo {author} {\bibfnamefont {Costas~M}\ \bibnamefont
  {Soukoulis}},\ }\href@noop {} {\emph {\bibinfo {title} {Wave propagation:
  from electrons to photonic crystals and left-handed materials}}}\ (\bibinfo
  {publisher} {Princeton University Press},\ \bibinfo {year}
  {2008})\BibitemShut {NoStop}%
\bibitem [{\citenamefont {Izrailev}\ and\ \citenamefont
  {Krokhin}(1999)}]{izrailev1999localization}%
  \BibitemOpen
  \bibfield  {author} {\bibinfo {author} {\bibfnamefont {FM}~\bibnamefont
  {Izrailev}}\ and\ \bibinfo {author} {\bibfnamefont {AA}~\bibnamefont
  {Krokhin}},\ }\bibfield  {title} {\enquote {\bibinfo {title} {Localization
  and the mobility edge in one-dimensional potentials with correlated
  disorder},}\ }\href@noop {} {\bibfield  {journal} {\bibinfo  {journal}
  {Physical review letters}\ }\textbf {\bibinfo {volume} {82}},\ \bibinfo
  {pages} {4062} (\bibinfo {year} {1999})}\BibitemShut {NoStop}%
\bibitem [{\citenamefont {Cheng}\ and\ \citenamefont
  {Song}(2012)}]{cheng2012fano}%
  \BibitemOpen
  \bibfield  {author} {\bibinfo {author} {\bibfnamefont {Mu-Tian}\ \bibnamefont
  {Cheng}}\ and\ \bibinfo {author} {\bibfnamefont {Yan-Yan}\ \bibnamefont
  {Song}},\ }\bibfield  {title} {\enquote {\bibinfo {title} {Fano resonance
  analysis in a pair of semiconductor quantum dots coupling to a metal
  nanowire},}\ }\href@noop {} {\bibfield  {journal} {\bibinfo  {journal}
  {Optics letters}\ }\textbf {\bibinfo {volume} {37}},\ \bibinfo {pages}
  {978--980} (\bibinfo {year} {2012})}\BibitemShut {NoStop}%
\bibitem [{\citenamefont {Pichler}\ \emph {et~al.}(2015)\citenamefont
  {Pichler}, \citenamefont {Ramos}, \citenamefont {Daley},\ and\ \citenamefont
  {Zoller}}]{pichler2015quantum}%
  \BibitemOpen
  \bibfield  {author} {\bibinfo {author} {\bibfnamefont {Hannes}\ \bibnamefont
  {Pichler}}, \bibinfo {author} {\bibfnamefont {Tom{\'a}s}\ \bibnamefont
  {Ramos}}, \bibinfo {author} {\bibfnamefont {Andrew~J}\ \bibnamefont {Daley}},
  \ and\ \bibinfo {author} {\bibfnamefont {Peter}\ \bibnamefont {Zoller}},\
  }\bibfield  {title} {\enquote {\bibinfo {title} {Quantum optics of chiral
  spin networks},}\ }\href@noop {} {\bibfield  {journal} {\bibinfo  {journal}
  {Physical Review A}\ }\textbf {\bibinfo {volume} {91}},\ \bibinfo {pages}
  {042116} (\bibinfo {year} {2015})}\BibitemShut {NoStop}%
\end{thebibliography}%
\end{document}